\documentclass{aa} %
\usepackage{CJK}
\usepackage{natbib}
\usepackage{graphicx}
\usepackage {amsmath,amssymb,eqnarray}
\usepackage{enumerate}
\usepackage{enumitem}
\usepackage{mathtools}
\usepackage[colorlinks, citecolor=blue, linkcolor=blue, urlcolor=blue]{hyperref}
\usepackage[flushleft]{threeparttable}

\begin{document}
\title{The XMM-SERVS X-ray eXtended Galaxy Cluster (XVXGC) catalog}
\author{Weiwei Xu\thanks{wwxu@pku.edu.cn; orcid:0000-0002-9587-6683}\inst{1,2,3} %[]
\and
Linhua Jiang\thanks{orcid:0000-0003-4176-6486}\inst{1,4}
\and
Ran Li\inst{2,3,5} 
\and
Bin Luo\inst{6,7}
\and
W. Nielsen Brandt\inst{8,9,10}
\and
Chaoli Zhang\inst{11}
\and
Thomas Erben\inst{12}
}

\institute{
Kavli Institute for Astronomy and Astrophysics (KIAA), Peking University, Beijing 100871, China%1
\and
National Astronomical Observatories (NAOC), Chinese Academy of Sciences, Beijing 100101, China%2
\and
Institute for Frontiers in Astronomy and Astrophysics, Beijing Normal University, Beijing 102206, China%3
\and
Department of Astronomy, School of Physics, Peking University, Beijing 100871, China%4
\and
School of Astronomy and Space Science, University of Chinese Academy of Science, Beijing 100049, China%5
\and
School of Astronomy and Space Science, Nanjing University, Nanjing, Jiangsu 210093, China%6
\and
Key Laboratory of Modern Astronomy and Astrophysics (Nanjing University), Ministry of Education, Nanjing, Jiangsu 210093, China%7
\and
%Collaborative Innovation Center of Modern Astronomy and Space Exploration, Nanjing, Jiangsu 210093, China%8
%\and
Department of Astronomy and Astrophysics, 525 Davey Lab, The Pennsylvania State University, University Park, PA 16802, USA%8
\and
Institute for Gravitation and the Cosmos, The Pennsylvania State University, University Park, PA 16802, USA%9
\and
Department of Physics, 104 Davey Lab, The Pennsylvania State University, University Park, PA 16802, USA%10
\and
College of Computer Science and Artificial Intelligence, Wenzhou University, 325035 Wenzhou, China%11
\and
Argelander-Institut f\"ur Astronomie (AIfA), Universit\"at Bonn, Auf dem H\"ugel 71, 53121 Bonn, Germany%12
}
\date{Received xx, 202x; accepted xx, 202x}

\abstract
%Context.
{To explain the well-known tension between cosmological parameter constraints obtained 
from the primary cosmic microwave background (CMB) and those drawn from 
galaxy cluster samples, we propose a possible explanation for the incompleteness of 
detected clusters are higher than estimated, i.e., certain types of galaxy groups or clusters have been overlooked in previous works.
}
%Aims.
{We aim to search for galaxy groups and clusters with particularly extended 
surface brightness distributions by creating a new X-ray-selected catalog of 
extended galaxy clusters from the XMM-Spitzer Extragalactic Representative 
Volume Survey (XMM-SERVS) data, based on a dedicated source detection and 
characterization algorithm that is optimized for extended sources.}
%Methods. 
{Our state-of-the-art algorithm is composed of wavelet filtering, 
source detection, and characterization. 
% based on extensive 
%simulations, we investigated the detection efficiency and sample purity. 
We make a visual inspection of the optical image, 
and spatial distribution of galaxies within the same redshift layer to confirm the existence of clusters and estimate the cluster redshift with the spectroscopic and photometric redshifts of galaxies. 
The growth curve analysis is used to characterize the detections.
%We also find our first ICM-detect clusters by cross-matching our detections with previous cluster catalogs. 
%investigate the detection efficiency and sample purity, 
}
%Results.
{We report a catalog of extended X-ray galaxy clusters detected from the XMM-SERVS data, 
named the XMM-SERVS X-ray eXtended Galaxy Cluster (XVXGC) catalog. It includes $141$ cluster candidates. Specifically, there are $52$ clusters previously identified as clusters with the intra-cluster medium (ICM) emission (class 3), $37$ ones previously known as optical or infrared clusters but detected as X-ray clusters for the first time (class 2), and $52$ identified as clusters for the first time (class 1). 
%Based on $200$ simulations, the contamination ratio of the detections that were identified as clusters by ICM emission and the detections that were identified as optical and infrared clusters in previous work is $xxx$ and $xxx$, respectively.   
Compared with the class 3 sample, the ``class 1 + class 2" sample is systematically fainter, and exhibits a flatter surface brightness profile. Specifically, the median flux in [$0.1-2.4$]~keV band for ``class 1 + class 2" and class 3 sample is $2.336\times 10^{-14}~$erg/s/cm$^2$ and $3.163\times 10^{-14}$~erg/s/cm$^2$, respectively. The median values of $\beta$ (the slope of cluster surface brightness profile) are $0.502$ and $0.577$ for the ``class 1 + class 2" and class 3 samples, respectively. This entire sample is available online together with the paper publication. 
}

\titlerunning{XVXGC clusters from XMM-SERVS data}
\authorrunning{W. Xu et al.}
\date{Received xx, 202x; accepted xx, 202x}

\maketitle

\section{Introduction}
\label{sec:introduction}

Galaxy clusters are the largest gravitationally bound systems in the universe, and are widely used to constrain cosmological models (e.g. \citealt{Bohringer2004, Vikhlinin2009, Mantz2010, Allen2011}), such as the constraints of components of dark matter, and the dark energy. 
Besides, galaxy clusters provide a dense environment to affect the galaxy's evolution. The statistical 
research of cluster member galaxies is vital to understanding the environmental effect in the formation
and evolution of galaxies (e.g., \citealt{Butcher1984, Lewis2002, Peng2010, Wang2020}). In any case, the identification of galaxy clusters is the basis of all cluster-based research. 

Since the Abell cluster catalog \citep{Abell1958} identified with the photographic plates, the identification of galaxy clusters has developed for more than 50 years, when more photometric data (e.g., \citealt{Wen2012}) and more accurate spectroscopic data (e.g., \citealt{Berlind2006}), even multi-bands data (e.g., \citealt{Wen2018, Bohringer2004, Planck2016a}) are used to make larger number and more accurate identification of galaxy clusters. With the N-body simulations or hydro-dynamical simulations, our knowledge of the baryon evolution increases rapidly (see references in \citealt{Borgani2011}).

Among all the identification methods, intra-cluster medium (ICM) emission traces
hot plasma inside the gravitational potential well, and 
provides a reliable tracer of massive galaxy clusters. In this method, the X-ray 
emission of the cluster comes mostly from the central area, and obeys less projection effect 
compared with other cluster tracers in other wavelengths, such as the Sunyaev-Zel'dovich effect (SZ effect, \citealt{Sunyaev1980}) of ICM in micro-wave and member galaxies in optical and Infra-red bands. What is more important, except for the dark matter, ICM comprises the most massive baryonic components of the cluster. There has been a large number of X-ray galaxy clusters 
identified from ROSAT (e.g., \citealt{Piffaretti2011}), XMM-Newton (e.g., \citealt{Pacaud2016}), 
and Chandra (e.g.,\citealt{Cavagnolo2009}). Besides X-ray of ICM, its SZ effect is another effective method to identify galaxy clusters, especially for high redshift clusters. 
%The SZ effect comes from the inverse Compton scattering of background cosmic microwave background (CMB) photons by the high energy ICM electrons. 
There are also a large number of galaxy 
clusters identified from the Planck survey \citep{Planck2016a}, 
the South Pole Telescope (SPT, \citealt{Bleem2015}), and the Atacama 
Cosmology Telescope (ACT, \citealt{Hilton2021}). 

In this paper, we take clusters identified in X-ray or micro-wave wavelength (with SZ effect) as ``ICM-detected clusters" because they are detected with the property of ICM. In addition, we take clusters identified in the optical or infrared (IR) band as ``OPT/IR clusters" because they are all detected with properties of member galaxies.

In this work, we aim to search for X-ray extended clusters with the XMM-SERVS survey data \citep{Chen2018, Ni2021} with the wavelet-based algorithm \citep{Pacaud2006, Xu2018}.
%There are xxx images provided, after the background? flare? noise? removal. 
The structure of the paper is as follows. 
Sec.~\ref{sec:data} describes the data briefly. 
Sec.~\ref{sec:method} presents the methodology. 
Sec.~\ref{sec:result} shows the cluster catalog and some discussion. 
In Sec.~\ref{sec:conclusion}, we conclude the project. 

\section{Data}
\label{sec:data}

In this work, we use the data products in the soft X-ray band of the XMM-Spitzer Extragalactic Representative Volume Survey (XMM-SERVS\footnote{\url{https://personal.science.psu.edu/wnb3/xmmservs/xmmservs.html}}, \citealt{Chen2018, Ni2021}). 
This survey covers $\sim13$~deg$^2$, comprised of XMM-Large-Scale Structure
(XMM-LSS, $5.3$~deg$^2$), Wide Chandra Deep Field South (W-CDF-S, 
$4.6$~deg$^2$), and ELAIS-S1 (ES1, $3.2$~deg$^2$) areas. These three contiguous 
fields are observed with three EPIC instruments of XMM-Newton (MOS1, MOS2, and PN). 
It has a comparably uniform X-ray coverage, 
with total flare-filtered exposure time of $2.7$~Ms, $1.8$~Ms, and $0.9$~Ms
respectively. 

The XMM-SERVS survey is designed to make detection of galaxy clusters with middle deep observations, filling the gap between deep pencil-beam and shallow large coverage X-ray surveys. 
For the XMM-LSS, W-CDF-S, and ES1 areas, the flux limits of X-ray point sources, are 
$1.7\times 10^{-15}$~erg/s/cm$^{2}$,
$1.9\times 10^{-15}$~erg/s/cm$^{2}$,
and $2.5\times 10^{-15}$~erg/s/cm$^{2}$, respectively,
in the $0.5–2.0$ keV band over $90\%$ of its area. The flux limit is calculated from the sensitivity map. As described in \cite{Ni2021}, the sensitivity map is calculated with the minimum source counts of detections minus the background, with the exposure time taken as weighted, as $S=(m-B)/t_{\rm exp}/{\rm EEF}/{\rm ECF}$, where the encircled energy fraction (EEF) and energy conversion factor (ECF) are also considered.

In the work of \cite{Chen2018} and \cite{Ni2021}, 
%there are $5242$, $4053$, $2630$ X-ray sources detected in total.
the X-ray images of XMM-SERVS survey are constructed after the screening for background flares, and the removal of events in the energy ranges overlapping with instrumental background lines, and mosaicked together.
%using the XMM-Newton Science Analysis System (SAS) task \textsc{evselect}. 
The exposure maps are vignetting-corrected. 
In our work, we use the mosaicked MOS1+MOS2+PN images in $0.5-2.0$~keV, including the event images, exposure maps, and background images.

There are multi-band resources in the field, such as XMM-Newton point-source catalog for the XMM-LSS Field \citep{Chen2018}, deep Hyper Suprime-Cam images and a forced photometry catalog in W-CDF-S \citep{Ni2019}, XMM-Newton point-source catalogs for the W-CDF-S and ELAIS-S1 Fields \citep{Ni2021}, the multi-band forced-photometry catalog in the ELAIS-S1 field \citep{Zou2021a}, photometric redshifts in the W-CDF-S and ELAIS-S1 Fields based on Infra-red forced photometry \citep{Zou2021b}. Combining the multi-band observations, there has been some research about galaxies and AGNs, such as the fitting of Spectral Energy Distributions of galaxies and source classification \citep{zou2022}, the selection and characterization of radio AGN \citep{Zhu2023}, identification and characterization of distant active dwarf galaxies \citep{Zou2023}, and uncover a sample of Compton-thick AGN as well as heavily obscured AGNs \citep{Yan2023}.

\section{Method}
\label{sec:method}

\subsection{Source detection}
\label{sec:detection}

Following the procedure in \cite{Pacaud2006} and \cite{Xu2018}, we run the wavelet filtering (\textsc{er$\_$wavelet}) on the X-ray event images in $0.5-2.0$~keV, taking the exposure map as the weight, and obtain the reconstructed X-ray image. 
In this step, the multi-resolution wavelet filtering is used to remove the Poisson noise, and the smooth reconstructed X-ray image is obtained for further analysis.
%The wavelet size range is set to 4 to 6 for the XMM-LSS and ES1 region, and 3 to 4 for the W-CDF-S region. The signal-to-noise ratio (SNR) threshold is set to $3.0\sigma$ for the XMM-LSS and ES1 region and $4.0\sigma$ for the W-CDF-S region. 
Then the \textsc{SExtractor} software \citep{Bertin1996} is used to detect the sources, where the exposure map is taken as the weight map. 
For the XMM-LSS, W-CDF-S, and ES1 region, there are $2869$, $2670$, $927$ sources detected respectively.
Finally, the maximum likelihood method is used to characterize detections. 

In the maximum likelihood fitting, we fit each detection with a point source model and a cluster model with C-statistics \citep{Cash1979}. 
The $\beta$-model is used for the surface-brightness profile of cluster, 
\begin{equation}
S_{\rm x}(r) \propto [1+(r/r_{\rm c})^2]^{-3\beta+0.5},
\label{eq:beta-model}
\end{equation} 
where $r_{\rm c}$ is the core radius of the cluster, and $\beta$ describes the slope of the brightness profile. The $\beta$ value of 2/3 is taken for a typical cluster.
The extension likelihood, EXT$\_$ML, is calculated as the difference between the detection likelihoods in the fitting of these two models. That is, EXT$\_$ML~=~$\rm{EXT\_DET\_ML-PNT\_DET\_ML}$. And the EXTENT is estimated as the core radius in the $\beta$-model. 

\begin{figure}
\centering
\includegraphics[width=0.45\textwidth]{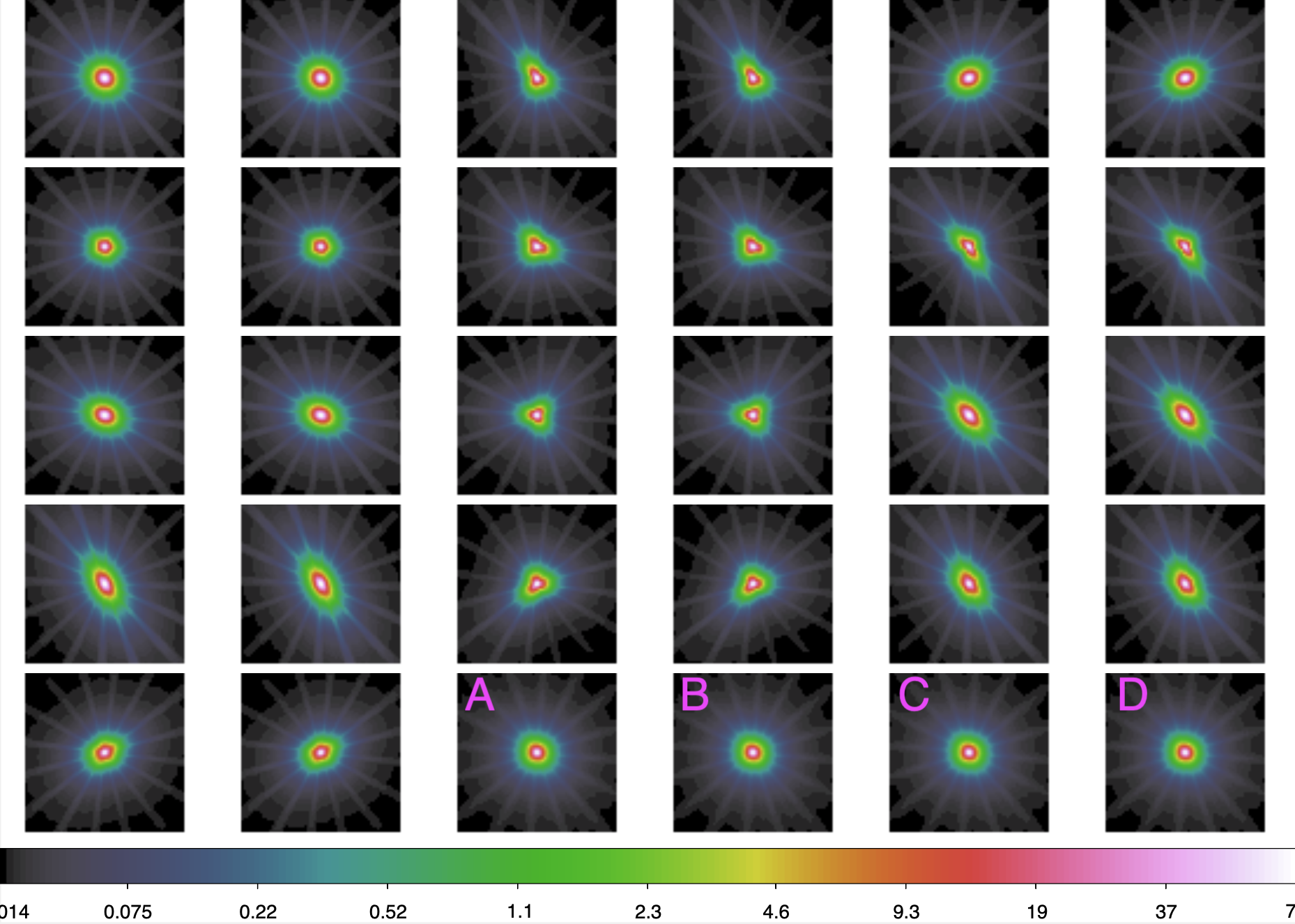}
\caption{Example of PSF reconstruction of one object, shown in logarithm scale. Panels are the PSF of each observation in the energy of $1$~keV and $2$~keV, except for the last four panels (panels A, B, C, D). Panels A and B are the combined PSF weighted with exposure map in $1$~keV and $2$~keV, separately. In panel C, the PSF image in $1$~keV and the PSF image in $2$~keV are combined. Panel D shows the combined PSF image multiplied by the source count, as the final PSF image.}
\label{fig:1_psf}
\end{figure}

In the point-source model-fitting, we need to reconstruct the point-spread function (PSF) image for each observation of each detection. It comes from the shape variation of PSF with the off-axis angle, instrument, and energy. For each detection, we use the XMM-Newton Science Analysis System (SAS) task \textsc{psfgen} to construct the PSF for its every observation in $1$~keV and $2$~keV observed by MOS1, MOS2, and PN, individually. Then all these PSF images for the detection are combined with an exposure map taken as a weight map. Fig.~\ref{fig:1_psf} shows an example of the PSF reconstruction step by step. All XMM-Newton observations in the XMM-SERVS survey are used.
%, except those with following obsID. For the XMM-LSS region, the removal ObsIDs are 0604280101, and 0677580101. For the W-CDF-S region, the removal obsID includes 0827210201, 0827212001, 0827221901, 0827230201, 0827230301, 0827230401, 0827230801, 0827231101, 0827231201, 0827231301, 0827231401, 0827231801, 0827232001. These observations are without P*IMAGE2000.FTZ and P*IMAGE3000.FTZ in their XMM Pipeline Products (PPS).}
%The used XMM-Newton observations are listed in Appendix~\ref{sec:list_obs}, $147$, $67$, and $31$ observations used in XMM-LSS, W-CDF-S, and ES1 area, respectively. %149-2, 80-13, 31

\begin{figure*}
\centering
\includegraphics[width=0.95\textwidth]{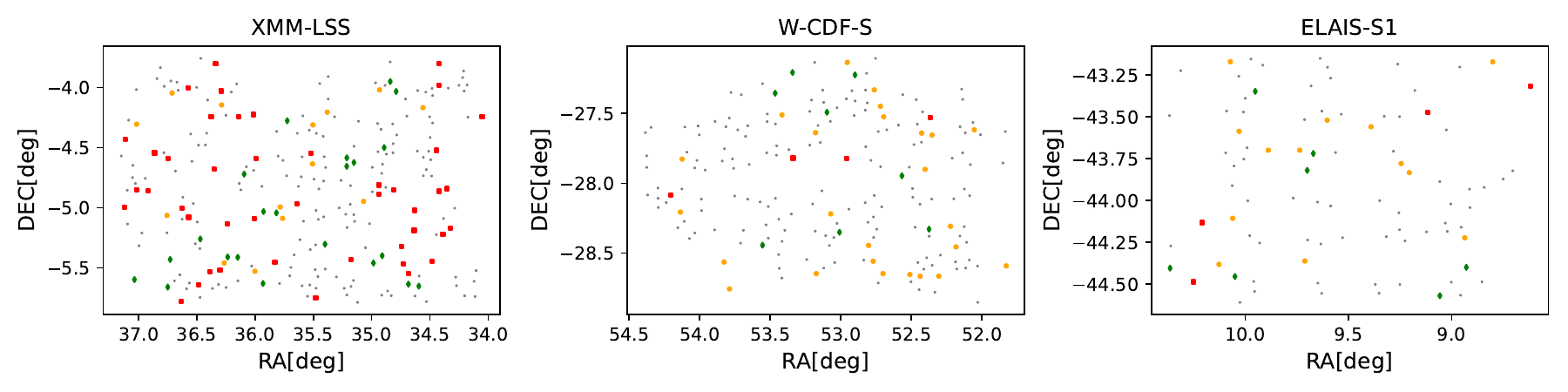}
\caption{The class 1, class 2, class 3, and false detections are shown in orange circles, green diamonds, red squares, and grey dots, respectively. The classification of detections is described in Sec.~\ref{sec:classification}. The three regions of XMM-SERVS are shown in sequence.}
\label{fig:radec}
\end{figure*}

\begin{figure}
\centering
\includegraphics[width=0.48\textwidth]{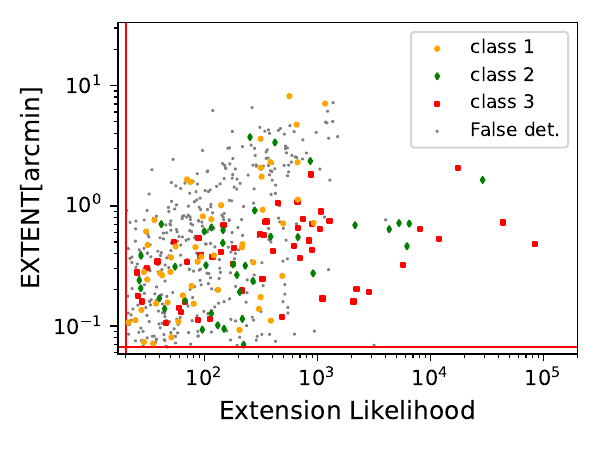}
\caption{The EXTENT-Extension Likelihood distribution of detections. The symbols are the same as in Fig.~\ref{fig:radec}. }
\label{fig:extent_extml}
\end{figure}

After that, thresholds are set for the detection likelihood, extension likelihood, EXTENT, and distance to the edge of survey coverage, as Eq.~\ref{eq:extent_criteria}, to select out extended sources. Finally, there are $241$, $187$, and $89$ extended detections in the XMM-LSS, W-CDF-S, and ES1 region, respectively. 
The positions of these extended sources are shown in Fig.~\ref{fig:radec}, and their distribution in EXTENT-Extension Likelihood parameter space is shown in Fig.~\ref{fig:extent_extml}. We refer to these extended detections as ``detections" in the later text.

\begin{equation}
\begin{split}
    \rm{EXT\_DET\_ML}>0,~~~~~~~~~~~~~~~~~~~~~~~~~~~~~~~~~~~~~~~~~~~~~~~~~\\
    \rm{PNT\_DET\_ML}>0,~~~~~~~~~~~~~~~~~~~~~~~~~~~~~~~~~~~~~~~~~~~~~~~~~\\
\rm{(EXT\_DET\_ML-PNT\_DET\_ML)=EXT\_ML} >20,~~~~~~\\
\rm{EXTENT}>4\arcsec,~~~~~~~~~~~~~~~~~~~~~~~~~~~~~~~~~~~~~~~~~~~~~~~~~~~~~~\\
\rm{distance\_to\_edge} >2.5\arcmin~~~~~~~~~~~~~~~~~~~~~~~~~~~~~~~~~~~~~~~~~~~~. 
\end{split}
\label{eq:extent_criteria}
\end{equation}

The wavelet-based algorithm selects the extended signals and discards the signals with small scales. This is both the advantage and disadvantage of this algorithm. It means we could identify more extended sources, instead of making a complete detection of clusters.

\subsection{Redshift estimation}
\label{subsec:redshift_estimation}

To validate the existence of an X-ray extended cluster, we need to find an over-density of galaxies at the same redshift layer. When the spatial distribution of this set of galaxies matches well with the X-ray contour, we take them as the cluster members and estimate the cluster redshift with their redshifts. 

The redshifts of galaxies are collected from the literature. 
The spectroscopic redshifts of galaxies are obtained from the Herschel Extragalactic Legacy Project (HELP), the Sloan Digital Sky Survey (SDSS) DR17, the VIMOS Public Extragalactic Redshift Survey (VIPERS), the Galaxy And Mass Assembly (GAMA), the Two Micron All Sky Survey Photometric Redshift catalog (2MPZ), 
PRIsm MUlti-object Survey (PRIMUS),
the VIMOS VLT Deep Survey (VVDS), the 6dF Galaxy Survey (6dFGS), the 2dF Galaxy Redshift Survey (2dFGRS), Carnegie-Spitzer-IMACS Survey (CSI), AAT Deep Extragalactic Legacy Survey (DEVILS), the UKIDSS Ultra-Deep Survey (UDSz), 3D-HST survey. 
The photometric redshifts of galaxies are obtained from the HELP, SDSS DR17, the Canada-France-Hawaii Telescope Legacy Survey (CFHTLS), the Dark Energy Spectroscopic Instrument (DESI), the Hubble Ultra Deep Field (UDF) catalog, Infrared Space Observatory (ISO), Dark Energy Survey (DES), Hyper Suprime-Cam (HSC) Deep Survey, Pan-STARRS1 Medium-Deep Survey (PS1MD), SWIRE optical imaging, VST Optical Imaging of CDF-S and ES1 (VOICE), SERVS DeepDrill survey. 
Besides, we gather galaxy redshifts from the NASA Extragalactic Database (NED\footnote{\url{https://ned.ipac.caltech.edu}}).  
The details and references are listed in Tab.~\ref{tab:g_catalogs}.

Galaxies with the offset to our detections $<1.5\arcmin$ are considered. 
To remove repetitive detections across surveys, we only reserve the information of one galaxy, when more than one galaxies with offset $<1\arcsec$ and $|\Delta z|<0.001$ with each other. The redshift priority decreases as the sequence of Tab.~\ref{tab:g_catalogs}. The galaxies from NED are given the lowest priority. 
The galaxies within the following areas are taken in XMM-SERVS:

\begin{equation}
\begin{split}
\rm{XMM-LSS: ~~RA}=33.5^{\circ}\sim37.6^{\circ}, ~~\rm{DEC}=-6.5^{\circ}\sim-3.0^{\circ},~~~\\
\rm{W-CDF-S: ~~RA}=51.0^{\circ}\sim55.5^{\circ}, ~~\rm{DEC}=-30.0^{\circ}\sim-26.5^{\circ},\\
\rm{ES1: ~~RA}=7.5^{\circ}\sim11.0^{\circ}, ~~~~\rm{DEC}=-46.0^{\circ}\sim-42.0^{\circ}.
\end{split}
\label{eq:region}
\end{equation}

\begin{table*}[t]
\caption{Overview of galaxy redshifts from literature and NED database$^*$.}
%ls_wc_es/cat_ls_wc_es/3get_norep_g.log
    \begin{threeparttable}
    \centering
        \begin{tabular}{c | c| c c c |c}
        \hline
        \hline
$z$      &                       Catalogs                 &    N(XMM-LSS)$^a$ & N(W-CDF-S)$^a$& N(ES1)$^a$     & Reference \\
        \hline
 $z$sp &     -  &           $0$ &   $29\,888$ & $10\,579$ & \cite{Zou2021b}\\
   &      HELP  &    $113\,192$ &   $48\,900$ & $17\,784$ & \cite{Shirley2021} \\
   &  SDSS DR17 &     $10\,893$ &         $0$ &       $0$ & \cite{Abdurrouf2021} \\
   &   DESI     &     $13\,642$ &    $1\,335$ &     $187$ & \cite{Zou2019}\\
   &        -   &     $42\,985$ &         $0$ &       $0$ & \cite{Chen2018}\\
   &   VIPER    &     $24\,963$ &         $0$ &       $0$ & \cite{Scodeggio2018}\\
   &    GAMA    &     $10\,108$ &         $0$ &       $0$ & \cite{Baldry2018}   \\
   &   2MPZ$^b$ &         $167$ &       $231$ &      $70$ & \cite{Bilicki2014}\\
   &   PRIMUS   &     $30\,939$ &   $17\,089$ &  $6\,954$ & \cite{Coil2011} \\
   &SXDF 100$\mu$Jy&      $315$ &         $0$ &       $0$ & \cite{Simpson2006,Simpson2012}\\ %simpson
   &     VVDS   &      $8\,472$ &    $1\,160$ &       $0$ & \cite{LeFevre2013} \\
   &    6dFGS   &         $139$ &        $53$ &      $79$ & \cite{Jones2004} \\
   &   2dFGRS   &           $0$ &    $1\,215$ &     $144$ & \cite{Colless2001} \\
   &    NED     &      $3\,589$ &    $2\,191$ &     $437$ & -\\
   N$_{\rm sum}$(zsp)$^c$& - & $135\,323$ & $55\,401$ &  $20\,172$ & -\\
% 113192+10893+13642+24963+10108+167+30939+315+8472+139+3589= 216419  
% 29888+48900+1335+231+17089+1160+53+1215+2191= 102062                
% 10579+17784+187+70+6954+79+144+437=36234                           
\hline
 $z$ph&      -  &           $0$ & $755\,714$& $806\,695$ & \cite{Zou2021b}\\
   &   HELP     & $4\,703\,183$ & $131\,296$& $928\,889$ & \cite{Shirley2021}  \\
   & SDSS DR17  &    $164\,521$ &       $0$ &        $0$ & \cite{Abdurrouf2021} \\
   &   DESI     &    $212\,720$ & $217\,615$& $146\,496$ & \cite{Zou2019}\\
   &        -   &    $347\,915$ &       $0$ &        $0$ & \cite{Chen2018}\\
   &  UDF       &           $0$ &  $9\,969$ &        $0$ & \cite{Rafelski2015}\\
   &  CFHTLS    & $1\,673\,935$ &       $0$ &        $0$ & \cite{Ilbert2006,Coupon2009}\\
   &  ISO       &           $0$ &       $0$ &      $147$ & \cite{LaFranca2004}\\ %es1
   &   NED      &     $25\,700$ & $18\,727$ &      $7\,966$ & -\\
N$_{\rm sum}$(zph)$^c$& -  &$7\,116\,131$& $1\,132\,723$ &$1\,888\,701$& -\\% 6660957, 928954, 1747233
% & \cite{Zou2021a} & 0 & 0 & n & DeepDrill, VIDEO, DES, ESIS, VOICE \\
% 4703183+164521+212720+1673935+25700= 6780059 
% 755714+131296+217615+9969+18727= 1133321 
% 806695+928889+146496+147+7966= 1890193   
  \hline
  \hline
         \end{tabular}
         \begin{tablenotes}
        \item[$^*$] The catalog is sorted with the decreasing sequence of the publication year and the priority. The spectroscopic and photometric redshifts are listed in the first and second parts separately. 
        \item[$^a$] The total number of galaxies in the region is counted in the survey coverage, as shown in Eq.~\ref{eq:region}.
        \item[$^b$] Only spectroscopic redshifts are used from the 2MPZ catalog.  
        \item[$^c$] The total number of galaxies after the removal of repetitive detections following the criteria as in Sec.~\ref{subsec:redshift_estimation}.
        \end{tablenotes}
    \end{threeparttable}
    \label{tab:g_catalogs}
\end{table*}

The estimation of cluster redshift is estimated in the following way. 
As the first step, we make the redshift histogram of galaxies within the range of $0-1.0$ and the bin size of $0.01$. Four peak bins including the largest number of galaxies are selected. For each of the four peak bins, galaxies with redshifts in the range of $z_{\rm peak}\pm0.015$ are taken as member candidates.
In \cite{Zou2021b}, the normalized photometric redshift error, $\Delta z_{\rm norm}=(z_{\rm phot}-z_{\rm spec})/(1+z_{\rm spec})$, has a median value of $-0.010$ for W-CDF-S sources, and $-0.013$ for ES1 sources. In this work, the $\pm 0.015$ is taken as a typical redshift range of member galaxies.
Then, the four peak bins are sorted with member numbers. 
The final average and standard deviation of the redshifts of member galaxies are obtained after the iterative $3\sigma$ clipping.
%(removal of the values out of 3 times the standard deviation around the mean redshift value) 
% iscipy.stats.sigmaclip
The standard deviation of redshift ($z$,err) is set as $0.005$ when it is smaller than $0.005$. 
This way, we obtain four candidates of the cluster redshift, as well as its error. 
The left and middle panels of the first row in Fig.~\ref{fig:visual_check_example} are examples of the spectroscopic and photometric redshift distribution. The four vertical lines label the locations of the four candidates of the cluster redshift.

To find out the final redshift of each detection, we obtain a reconstructed X-ray image, as described in Sec.~\ref{sec:detection}, as well as the corresponding X-ray contour. We also obtain the $r$-band image from DES.
For each of the four candidate redshifts, the position of galaxies within 
$z \pm (3\times z$,err) together with the X-ray contour are overlaid over the $r$-band image and the reconstructed X-ray image. 
This procedure is undertaken for spectroscopic redshifts and photometric redshifts of galaxies respectively. 
The left and middle panels of the second row in Fig.~\ref{fig:visual_check_example} are shown as an example. The galaxies whose redshifts within $\pm 0.015$ around the four candidates of cluster redshift values are labeled with symbols.

In addition, the RGB image combining the $g, r, i$ bands is downloaded from the DES survey\footnote{\url{https://des.ncsa.illinois.edu/desaccess/home}}.
%(example link 
%here\footnote{\url{http://skyserver.sdss.org/dr16/SkyServerWS/ImgCutout/getjpeg?ra=34.051\&dec=-4.241\&scale=0.5\&height=600\&width=600}}) for SDSS,
The RGB image helps to point out galaxies with similar colors and check whether the galaxies in the specific redshift layer have consistent spatial distribution as the X-ray contour. An example of RGB image is shown as the left panel of the third row in Fig.~\ref{fig:visual_check_example}.

By visual check, as shown in Sec.~\ref{sec:visualcheck}, we select the cluster redshift out of the four photometric redshift candidate values, four spectroscopic redshift candidate values, and the redshift of the bright central galaxy. The redshift source information is listed in the column ``$z$,src" of the final catalog.
When we obtain the redshift and its error of the cluster, galaxies with the redshift within $z \pm (3\times z$,err) are taken as cluster members. The number of members is also listed in the final catalog. 
In Fig.~\ref{fig:visual_check_example}, an example of the redshift histogram and spatial distribution of member galaxies are shown as the right panel in the first row and the right panel in the second row.

\subsection{Visual check for genuine clusters}
\label{sec:visualcheck}

We make visual checks of detections to separate genuine clusters and false detections. We consider the position of members, the position of cluster detections, $r$-band image overlaid with X-ray contours, and the reconstructed X-ray image overlaid with X-ray contours, as well as the redshift histogram of members. The detections with the following features are taken as genuine clusters:
\begin{itemize}
    \item The X-ray contour has a regular or symmetric shape;
    \item The X-ray emission has a strong signal-to-noise ratio;
    \item The spatial distribution of members matches well with the X-ray contour, and one or several bright members are located at the X-ray peak;
    \item Members have similar colors in the RGB image;
    \item The redshift histogram of members has a Gaussian$-$like distribution.
\end{itemize}

The detections without these listed features are taken as false detections, and discarded in our further discussion. The false detections comprise detections with weak X-ray emission, X-ray emission with irregular contour, and X-ray emission with small size, without obvious central bright galaxies in similar RGB colors within the cluster redshift range, without obvious overdensity of galaxies in the cluster redshift range.

For the reliability of detected genuine clusters, we discard also some plausible and ambiguous detections with less credibility. Probably, there are some clusters misclassified as false detections. Thus, the final cluster catalog in this work includes some newly identified clusters with comparable reliable signals, instead of being a complete cluster catalog.

\begin{figure*}
    \centering
    \includegraphics[height=0.24\textwidth]{z_hist_es_706_zg_part.pdf}
    \includegraphics[height=0.24\textwidth]{z_hist_es_706_zg_ph_part.pdf}
    \includegraphics[height=0.24\textwidth]{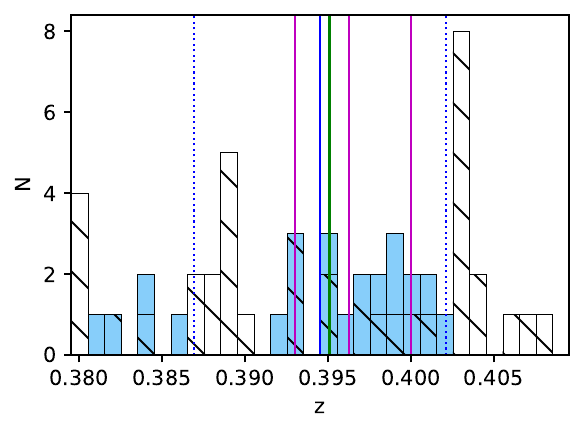}
    \includegraphics[height=0.27\textwidth]{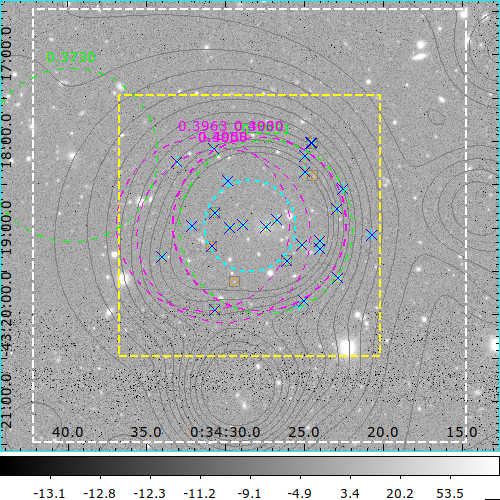}
    \includegraphics[height=0.27\textwidth]{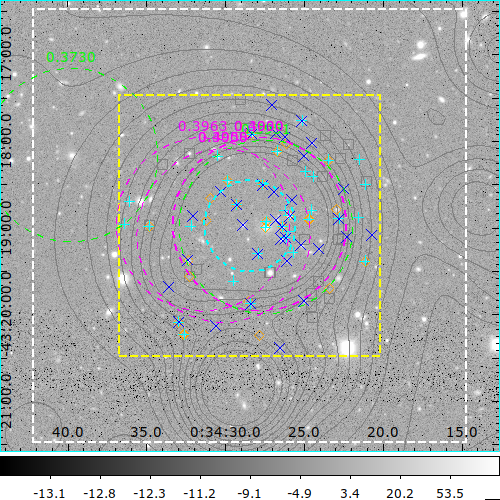}
    \includegraphics[height=0.27\textwidth]{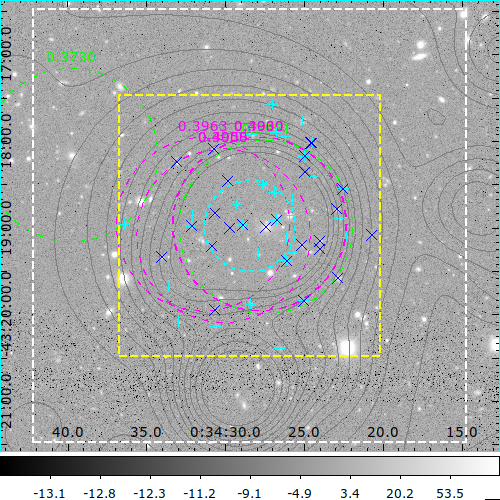}
    \includegraphics[height=0.18\textwidth]{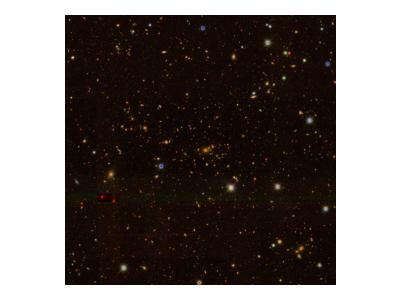}
    \includegraphics[height=0.18\textwidth]{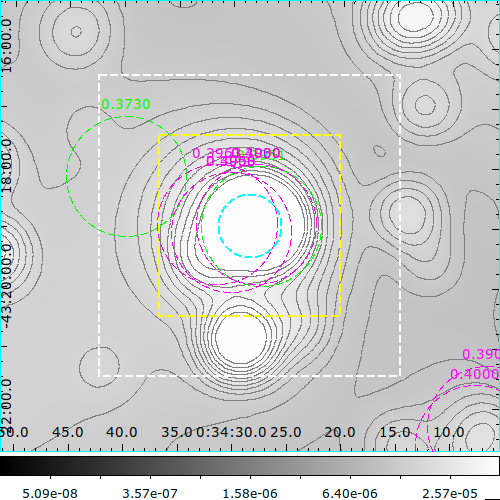}
    \includegraphics[height=0.18\textwidth]{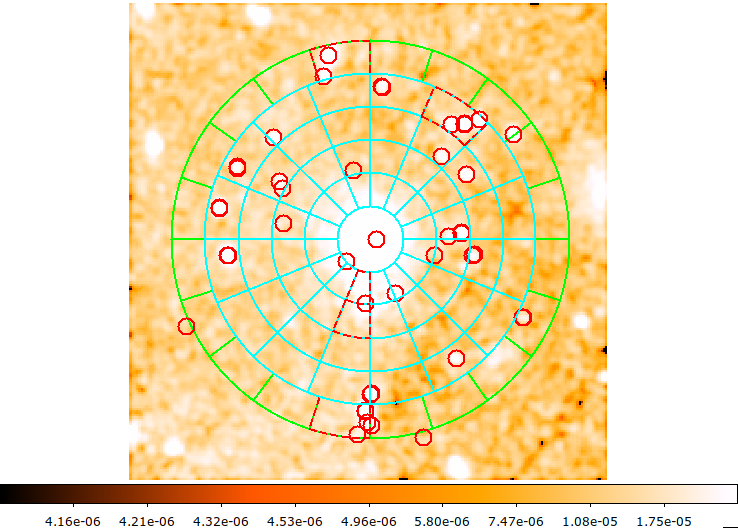}
    \includegraphics[height=0.15\textwidth]{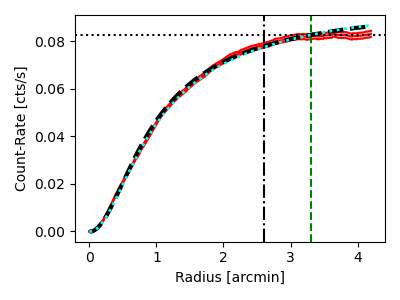}
    \caption{Images and plots of the cluster ``es$\_$1" for the visual check, as an example. 
    \textbf{[First row].}
    The three panels show the redshift distribution of galaxies within the offset $<1.5\arcmin$. 
    \textbf{The left panel} shows the spectroscopic redshift histogram of galaxies. The redshift range is set as $0-1.0$. The bin width is $0.01$. The blue, cyan, orange, and grey vertical line shows the location of the four highest bins. The magenta and green histograms are for the redshift of previous ICM-detected clusters and OPT/IR clusters with offset $<3.0\arcmin$ respectively. 
    \textbf{The middle panel} is the same as the left panel, but for photometric redshifts.
    \textbf{The right panel} is the redshift histogram of member galaxies, blue bins for spectroscopic redshifts, and black hatched bins for photometric redshifts. The redshifts within $z_{\rm peak}\pm0.015$ are plotted with 30 bins. The blue solid and blue dotted vertical lines show the cluster redshift and its 1$\sigma$ range. The magenta and green vertical lines are the redshifts of previous ICM-detected clusters and OPT/IR clusters.
    \textbf{[Second row].}
    The three panels show the spatial distribution of galaxies in a specific redshift layer corresponding to the three panels of the first row, in sequence. 
    \textbf{The left panel} is the $r$-band image in log scale, overlaid with galaxies redshifts within $\pm 0.015$ around the four highest redshift bins (blue ``x", cyan ``+", orange diamond, and grey box in sequence), X-ray contour in grey, location of previously identified clusters with redshift values (referring to Sec.~\ref{sec:classification}, magenta circle for ICM-detected cluster, green circle for OPT/IR cluster, radius as the smaller value of $0.5$~Mpc and $1.0\arcmin$), the location of XVXGC cluster (cyan circle with a radius of ``EXTENT" value), the white box with the size of $5\arcmin\times 5\arcmin$, the yellow box with the size of $3.0\arcmin\times 3.0\arcmin$.  %the black box with the size of $10\arcmin\times 10\arcmin$, 
    \textbf{The middle panel} is the same as the left panel, but for the photometric redshifts. 
    \textbf{The right panel} is the same as the left panel, but overlaid only the location of member galaxies (blue ``x" for spectroscopic redshift, cyan ``+" for photometric redshift).
    \textbf{[Last row].} 
    \textbf{The left panel} shows the RGB image in the size of $10\arcmin\times10\arcmin$. 
    \textbf{The left-middle panel} is the reconstructed X-ray image in log scale, overlaid with the X-ray contour in grey, and previously identified clusters whose symbols are consistent with panels in the second row.
    \textbf{The right-middle panel} is the reconstructed X-ray image in log scale, overlaid with the source regions (cyan sectors) and background regions (green sectors) for the growth curve analysis. The red circles outside the central cyan circle, and red dashed sectors are masks for contaminates. 
    \textbf{The right panel} is the growth curve plot, showing the integrated count rate versus the radius in red ($1~\sigma$ range in pink), where black dot-dashed and green dashed vertical lines label the $R_{500}$ and the significant radius, the horizontal dotted line shows the count rate in the plateau. The black curve is the best-fitting model for the growth curve, with the $\beta$-value labeled in the legend. The model with typical $\beta=2/3$ is plotted in a cyan dotted curve as a comparison. }
    \label{fig:visual_check_example}
\end{figure*}

In Fig.~\ref{fig:visual_check_example}, we show an example of the figures and plots for the visual check for one detection. Although the locations and redshifts of previously identified clusters are overlaid, this information is only used to classify genuine cluster detections, as described in Sec.~\ref{sec:classification}, and not taken into account in this section to separate genuine clusters and false detections. 

\subsection{Characterization of genuine clusters}

After the source detection, redshift estimation, and removal of false detection, we estimate the physical parameters of the cluster candidates, including the size, flux, luminosity, mass, and slope of the surface brightness profile.

\subsubsection{Growth curve analysis} 
\label{subsec:growth_curve}

The growth curve analysis is made to characterize cluster detections. The integrated count rate within the radius of $r_{\rm src}$ is plotted versus the radius, after the contribution of background and contaminates are removed. An example of the region selection and the growth curve are shown in the last two panels of Fig.~\ref{fig:visual_check_example}.

%As the first step, 
%we mask off the region which is refilled with nearby median? the count rate of nearby regions. The reason is the refilled image in this region is not necessarily a good indication of the emission from the central source. 
%Secondly, 

Firstly, we mask off contamination from the X-ray sources listed in the last part of Tab.~\ref{tab:gc_catalogs} using circle masks with a radius of $15\arcsec$. Some weak or compact X-ray clusters might be identified as X-ray sources, especially clusters with central AGN, therefore we do not add any masks at the central area with the radius of $20\%~r_{\rm src}$. 
%The X-ray point-source catalogs and the quasar catalog listed in Tab.~\ref{tab:catalogs} are considered. 

The background is estimated with the average count rate in the annulus with radius from $r_{\rm src}$ to $r_{\rm bgd}$. 
We separate the background area into 20 sectors and correct the count rate in each sector with the ratio of unmasked pixels. This way, we obtain the median and standard deviation of the total count rate for background sectors. The background sector deviated from the median with $>2.3\sigma$ is removed. The final estimation of the background count rate is calculated. 

In the source area, the region with a radius of $20\%~r_{\rm src}$ to $r_{\rm src}$ is separated into 4 annuli, and further divided into 16 sectors each. In each annulus, sectors with contaminates are removed with the same procedure as the background. 
In the central area with a radius of $20\%~r_{\rm src}$, emission from all pixels is considered in the growth curve. 
Finally, we obtain the integrated count rate for a set of radii after the removal of the background. 

%And the sum of count rate is calculated within each sector, and corrected with the unmasked number ratio of pixels. Then, we mask off sectors with the sum of the count rate higher or lower than the median value in the same annulus by more than 2.3 times the standard deviation. Finally, we obtain the integrated count rate within some specific radius, and plot the growth curve. Here, the correction with the ratio of good sectors is also considered. 

In the process of the growth curve, the $r_{\rm src}$ and $r_{\rm bgd}$ are usually taken as $5.0\arcmin$ and $6.0\arcmin$, respectively.
For some complex systems, the values of $r_{\rm src}$ and $r_{\rm bgd}$ are changed manually to obtain a steady growth curve with a flat plateau outside. For some clusters, some left contaminates in the source area cause a bad performance of the growth curve. Thus, we mask off some more sectors manually with contaminates identified by a visual check.
%The determination is made by looking through the multiple images, as well as the position and the redshift value of galaxies in the nearby region, as shown in Fig.~\ref{fig:example}. 

As shown in the right panel of the last row in Fig.~\ref{fig:visual_check_example}, in the growth curve, the integrated count rate increases with the radius and turns into a plateau at the significant radius. Outside the significant radius, the X-ray contribution from the central source is negligible compared with the background fluctuation. Thus, we obtain the the significant radius ($R_{\rm sig}$) as the inner radius of the plateau area, and the count rate in the plateau ($CR_{\rm sig}$). These two parameters are used to characterize our clusters, as described in Sec.~\ref{sec:flm}.

\subsubsection{Flux, luminosity and mass estimation}
\label{sec:flm}

In this subsection, we make an estimation of flux, luminosity, and mass for each cluster, with the value of significant radius and the count rate in the plateau from the growth curve analysis. The procedure is described in detail in \cite{Xu2022}. We list only the main steps below.

Firstly, we assume $R_{500}=R_{\rm sig}$, and obtain the total mass 
within $R_{500}$ ($M_{500}$). Then, we derive the bolometric X-ray 
luminosity ($L_{\rm x}$) and temperature ($T_{\rm x}$), using scaling relations from the Eq.~$23$-$26$ of \citet{Reichert2011}.
With the APEC model, we further obtain the luminosity within $R_{500}$ in [$0.1-2.4$] keV band ($L_{500}$), as well as the flux in the band ($F_{500}$). 
After the conversion factor between the count rate and flux is obtained for the instruments of MOS1+MOS2+PN with PIMMS \citep{Mukai1993}, we derive the total count rate within $R_{500}$ ($CR_{500}$). In this step, 
the absorption of neutral hydrogen is obtained from the HI4PI survey \citep{HI4PI}.

Assuming the typical $\beta$-profile (Eq.~\ref{eq:beta-model}) with $\beta=2/3$ and take $CR_{\rm 500}$ as $CR_{\rm sig, est}$, we iteratively repeated above steps until $CR_{\rm sig,est}$ = $CR_{\rm sig}$.
This way, we obtain $CR_{500}$ and $R_{500}$, as well as $M_{500}$, $L_{500}$, $T_{\rm x}$, and $F_{500}$. 
The $1\sigma$ error of $CR_{\rm sig}$ is estimated with the count rate uncertainty in the plateau. Then, the error of $CR_{\rm 500}$, $F_{\rm 500}$,  $L_{\rm 500}$, $T_{\rm x}$, $M_{\rm 500}$, and $R_{\rm 500}$ are estimated in sequence.

\subsubsection{$\beta$-value from the growth curve}

As mentioned in Sec.~\ref{sec:detection}, the typical surface-brightness profile of cluster can be described with $\beta$-model (Eq.~\ref{eq:beta-model}), and the $\beta$ value reflects the steepness of the profile. The larger $\beta$ value, the steeper profile. In this section, we estimate $\beta$ value for each detection with the growth curve. 

The PSF of XMM-Newton EPIC instruments varies largely with the energy, off-axis angle, and instrument. 
\cite{Read2011} provides the fully 2-D characterization of the PSF as a function of energy, off-axis angle, for each EPIC instrument. 
The PSF of PN can be described using a 2-D King profile ($B(r)$, Eq.~\ref{eq:2d-king}), while a further 2-D Gaussian function ($G(r)$, Eq.~\ref{eq:2d-gaussian}) is needed for MOS1 and MOS2 for the excess emission at the core. 
In the 2-D King profile, the $r_0$ is the core radius, $\alpha$ is the power-law slope, $\epsilon$ is the ellipticity, and $\theta$ is the angle of ellipticity. In the 2-D Gaussian function, the FWHM is the full width at half maximum, Norm is the normalization ratio of the Gaussian peak to the King peak. 

Some other features are coming from the support structure features, including the radially-dependent primary and secondary spoke structures, and the large-scale azimuthal modulation. However, the X-ray data used in this work is mosaicked from multiple observations of all three EPIC instruments. Thus, it is difficult to model these spatial-dependent features.

\begin{equation}
\begin{split}
B(r)=\frac{A}{[1+(r/r_0)^2]^\alpha},~~~~~~~~~~~~~~~~~~~~~~~~~~~~~~~~~~~~~~~~\\
\\
r(x,y,\theta)=\sqrt{[(x~{\rm cos}\theta+y~{\rm sin}\theta)^2]+\frac{[(y~{\rm cos}\theta-x~{\rm sin}\theta)^2]}{(1-\epsilon)^2}}.
\end{split}
\label{eq:2d-king}
\end{equation}

\begin{equation}
    G(r)={\rm Norm}\cdot A~e^{-4~{\rm ln}(2)(r/{\rm FWHM})^2}.
    \label{eq:2d-gaussian}
\end{equation}

As described in \cite{Read2011}, the PSF parameters for the MOS1, MOS2, and PN are provided in the ELLBETA parameters of the files, XRT1$\_$XPSF$\_$0016.CCF, XRT2$\_$XPSF$\_$0016.CCF, XRT3$\_$XPSF$\_$0018.CCF. 
The best sets of PSF parameters are provided in 8 energy values (0.1 keV, 1.5 keV, 2.75 keV, 4.25 keV, 6 keV, 8 keV, 10.25 keV, 15 keV) and 7 off-axis angles ($0\arcmin$, $1\arcmin$, $3\arcmin$, $6\arcmin$, $9\arcmin$, $12\arcmin$, $15\arcmin$). 
The off-axis angle of $1\arcmin$ is not provided for MOS1. % mos2,3; mos1 has no 1'
The PSF parameters in 1.5~keV with the off-axis angle of $9\arcmin$ are taken as the representative PSF of our data. 

There are 5 parameters in the PSF model, $r_0$, $\alpha$, $\epsilon$ of the 2-D King model, FWHM and Norm of the 2-D Gaussian model. The $\theta$ in the 2-D King model is always 0.
The values of $r_0$ and $\alpha$ of MOS1, MOS2, and PN, are averaged to obtain their values of final PSF.
There is no asymmetric information included in the growth curve, thus the ellipticity is set to 0.
However, the ellipticity will systematically enlarge the PSF size in mosaicked images, and might bring some bias to our estimation of $\beta$-model parameters in some way. 
%However, the ellipticity will definitely enlarge the core radius of PSF in mosaic images, thus we further divide the $r_0$ by ellipticity. 
Because only MOS1 and MOS2 have the 2-D Gaussian component, 
we take the average FWHM of MOS1 and MOS2 as the final FWHM parameter and divide the sum of their Norm values by 3 as the final Norm parameter. The final set of PSF parameters are $(r_0, \alpha, \epsilon, {\rm FWHM}, {\rm Norm})=(7.605\arcsec, 1.624, 0, 4.304\arcsec, 0.582)$.

In the fitting of the growth curve, 
the $\beta$-model (Eq.~\ref{eq:beta-model}) is convolved with the PSF model. The parameters are estimated with the Markov Chain Monte Carlo (MCMC) fitting 
using the $EMCEE$ package\footnote{\url{https://emcee. readthedocs.io/en/stable}} \citep{emcee}. 
The $\beta$-value is constrained within the range of $0.3-1.0$. 
We use $50$ chains with the original length of $5\,000$ steps each, and discard the first $2\,000$ steps. 

The last panel of Fig.~\ref{fig:hist_para} shows the posterior distribution of $\beta$-value. 
Although there are peaks at the high and low-value end representing ultra-steep or ultra-flat surface brightness profiles, the newly identified X-ray clusters systematically have lower $\beta$-value than previously identified X-ray clusters. Their flat X-ray profile might be a reason for the incomplete detection of X-ray clusters. This tendency is consistent with the RXGCC sample. 

\subsection{Classification of genuine clusters}
\label{sec:classification}

After the detection and characterization, we check whether cluster candidates were detected previously.
We refer to both X-ray clusters and SZ clusters as ``ICM-detected clusters", and refer to both optical and infrared clusters as ``OPT/IR clusters". 

We collect literature and NED database for previously identified clusters, listed in Tab.~\ref{tab:gc_catalogs}.
For convenience, we refer to clusters from literature or NED as ``clusters from the literature" in later text.
In the NED database, we take systems with the following types as clusters, 
\texttt{cluster of galaxies, group of galaxies}. 
Out of the NED clusters, systems with names beginning with the following are taken as ICM-detected clusters, ``3XLSS", ``ACT-CL", ``ECDF-S", ``RCC", ``RzCS", ``SMACS", ``SPT-CL", ``SXDF", ``X-CLASS", ``XLSS", ``XLSSC", ``XLSSsC", ``XMM-LSS", ``XMMU", ``XMMXCS", ``XXL-N". 
Otherwise, the systems are taken as OPT/IR clusters, whose names begin with
``400d", ``ABELL", ``CCPC-z", ``CDFS:[AMI2005]", ``CDGS", ``CFHT-D CL", ``CFHT-W CL", ``CFHTLS CL", ``CFHTLS c", ``CFHTLS:[DAC2011] W1", ``CFHTLS:[SMD2018a] W1", ``CL", ``CVB", ``ClG", ``G3Cv10", ``HSCS", ``JKCS", ``LCLG", ``MZ", ``PGC1", ``RCS1", ``redMaPPer", ``RM", ``SCG", ``SL", ``SWIRE CL", ``SXDS:[MHE2007]", ``SpARCS", ``UDSC", ``WHL", ``[AMP2011]", ``[DJ2014]", ``[DRV2022]", ``[HMC2016]", ``[LIK2015]", ``[LMR2016]", ``[MOH2018]", ``[MSP2015]", ``[PCG2016]", ``[SCP2009]", ``[TKO2016]", ``[VCB2006]", ``[WH2018]", ``[YHW2022]".
%``AM", %``APMUKS(BJ)", %``ESO", %``GALEXASC", %``[DLT2009]", 

\begin{table*}[t]
\caption{Overview of identified clusters from literature and NED database, and X-ray sources identified in ROSAT and XMM-Newton.}
%ls_wc_es/cat_ls_wc_es/1catgc_servs.log
    \begin{threeparttable}
    \centering
    \tiny
        \begin{tabular}{c | c| c c c | c c}
        \hline
        \hline
    Catalogs        &  Number   & N(XMM-LSS)$^1$    & N(W-CDF-S)$^1$    &N(ES1)$^1$& Survey    & Reference  \\
        \hline
    X-ray cluster   & $12\,247$ & $0$               & $26$              & $4$   &  eROSITA                  & \citealt{Bulbul2024}\\
                    &     $944$ & $0$               & $0$               & $0$   &  ROSAT                    & \citealt{Xu2022}\\
                    & $1\,559$  & $34$              & $8$               & $6$   &                           & \citealt{Koulouridis2021}  \\
                    & $10\,382$ & $11$              & $0$               & $0$   &   ROSAT                   & \citealt{Finoguenov2020} \\
                    & $302$     & $92$              & $0$               & $0$   &-                          & \citealt{Adami2018} \\
                    & $1\,490$  & $5$               & $0$               & $1$   &-                          & \citealt{Wen2018}\\
                    %   & $520$ & $0$               & $0$               & $0$   & \citealt{Clerc2016} \\ %catCluster$-$SPIDERS$\_$RASS$\_$CLUS$-$V2.0.fits, out of CODEX 
                    &    $107$  & $26$              & $0$               & $0$   & XMM-Newton XXL       &\citealt{Pacaud2016,Pierre2016}\\% XLSSC %XXL2, same with Pierre2016
                    &      $46$ & $0$               & $46$              & $0$   & XMM-Newton, Chandra  & \citealt{Finoguenov2015}\\
                    % & $100$   & $17$              & $0$               & $0$   &   &\citealt{Pierre2016}\\ %XXL1
                    & $904$     & $0$               & $0$               & $0$   &  XMM-Newton, SDSS    & \citealt{Takey2011,Takey2013,Takey2014,Takey2016} \\%175+185+345+145+54, 0,0,0
                    % & $2\,740$& $2\,740$          & $0$               & $0$   & ROSAT & \citealt{Clerc2020,Kirkpatrick2021} \\
                    & $503$     & $32$              & $0$               & $5$   & XMM-Newton          &\citealt{Mehrtens2012}  \\ %XCS
                    & $422$     & $14$              & $1$               & $3$   &XMM-Newton XXL       & \citealt{Clerc2012}   \\  %XCLASS 
                    & $1\,743$  & $1$               & $4$               & $0$   &ROSAT                      &\citealt{Piffaretti2011}\\ %MCXC
                    & $57$      & $57$              & $0$               & $0$   & XMM-Newton XXL      & \citealt{Finoguenov2010} \\ 
                    & $242$     & $0$               & $4$               & $0$   &                           & \citealt{Burenin2007}  \\
                    % &   $263$ & $0$               & $0$               & $0$   & Swift&\citealt{Liu2015} \\
                    &    $579$  & $0$               & $0$               & $0$   & ROSAT                     &\citealt{Ledlow2003} \\
                    &    $283$  & $0$               & $0$               & $0$   & ROSAT                     &\citealt{Ebeling1996} \\ %XBACS
                    & -         & $85$              &$22$               & $1$   & NED                       & - \\
    %                & -         & $73$              &$17$               & $1$   & NED                       & - \\
    N$_{\rm sum}$   &    -      & $357$             & $111$             & $20$  & -                         & -         \\
        \hline %34+11+92+5+26+32+14+1+57+85=357; 26+8+46+4+1+4+22=111; 4+6+1+5+3+1=20
SZ cluster          & $4\,195$  & $10$              &$1$                & $6$   & ACT                       &\citealt{Hilton2021}         \\ 
                    %& $419$    & $0$               &$0$                & $3$ ? &{\it Planck}, SPT          &\citealt{Melin2021}        \\ %J/A+A/647/A106/catalog
(detected in Microwave)&$2\,323$& $1$               &$0$                & $2$   & Planck, RASS         &\citealt{Tarrio2019}      \\
                    & $225$     & $0$               &$0$                & $2$   & Planck, RASS         &\citealt{Tarrio2018}      \\
                    & $182$     & $8$               &$0$                & $0$   &ACT                        &\citealt{Hilton2018}       \\
                    %& $89$     & $0$               &$0$                & $1$?  &  SPT                      &\citealt{Bayliss2017}     \\ %J/ApJ/837/88/table2
                    & $1\,653$  & $1$               &$0$                & $1$   & Planck               &\citealt{Planck2016a}      \\%PSZ2
                    & $1\,227$  & $1$               &$0$                & $1$   & Planck               &\citealt{Planck2014,Planck2015} \\
                    & $189$     & $0$               &$0$                & $0$   & Planck               &\citealt{Planck2011}       \\
                    & $677$     & $0$               &$0$                & $4$   & SPT                       &\citealt{Bleem2015}  \\  %SPT
                    %& $75$     & $0$               &$0$                & $1$?  & SPT                       &\citealt{Ruel2014}        \\ % J/ApJ/792/45/table1
                    %& $224$    & $0$               &$0$                & $0$   & SPT                       &\citealt{Reichardt2013}    \\
                    & $91$      & $0$               &$0$                & $0$   & ACT                       &\citealt{Hasselfield2013} \\
                    & $23$      & $0$               &$0$                & $0$   & ACT                       &\citealt{Marriage2011}     \\
                    & -         & $0$               &$0$                & $4$   & NED                       & - \\
   N$_{\rm sum}$    & -         & $21$              &$1$                & $20$  & -                         & - \\
        \hline %10+1+8+1+1=21; 6+2+2+1+1+4+4=20
%& \citealt{Yang2021} & ? &  & $?$ &$?$ & $?$ & DESI \\
OPT/IR cluster       &  $189$ & $65$     &$65$                & $50$   & Spitzer                       & \citealt{Gully2024}           \\
(detected by members)& $1\,921$ & $135$            &$0$                & $0$   & HSC                       & \citealt{Oguri2018}           \\
                    & $19$      & $0$              &$19$               & $0$   & MUSYC, ACES               & \citealt{Dehghan2014}  \\ % find 13 galaxy clusters and 6 large groups/small clusters.
                    & $71\,743$ & $76$              &$0$                & $0$   & SDSS                      &\citealt{Oguri2014}   \\
                    & $26\,898$ & $56$              &$18$               & $14$  & SDSS                      &\citealt{Rykoff2016}          \\  %26111+787
                    & $25\,325$ & $36$              &$0$                & $0$   & SDSS                      &\citealt{Rykoff2014}          \\
                    & $47\,600$ & $28$              &$31$               & $25$  & 2MASS, WISE, SuperCOSMOS  &\citealt{Wen2018}              \\ %know_21475(26,5,4), new_26125 (2,26,21)
                    & $25\,419$ & $25$              &$0$                & $0$   & SDSS                      &\citealt{Wen2015}              \\ 
                    &  $2\,092$ & $4$               &$0$                & $0$   & SDSS                      &\citealt{Wen2013}              \\ 
                    &$132\,684$ & $150$             &$0$                & $0$   & SDSS                      &\citealt{Wen2009,Wen2012}     \\ 
                    %& $2\,433$ & $0$               & $0$               & $0$   & WISE                      &\citealt{Gonzalez2019}       \\
                    % & ? &  & $?$ &$?$ & $?$ & ? & \citealt{Yang2007} \\
                    & $13\,340$ & $2$               &$2$                & $2$   & -                         &\citealt{Abell1958,Abell1989}  \\
                %    & -         & $455$             &$70$               & $14$  & NED                       & - \\
                    & -         & $577$             &$94$               & $21$  & NED                       & - \\
   N$_{\rm sum}$    & -         & $1\,154$          &$229$              & $112$  & -                         & - \\
        \hline %65+135+76+56+36+28+25+4+150+2+577=1154;  65+19+18+31+2+94=229; 50+14+25+2+21=63;
X-ray source        & $6\,683$  & $0$               &$4\,053$           &$2\,630$& XMM-Newton          &\citealt{Ni2021}           \\%wc_4053, es_2630
                    & $5\,242$  & $5\,242$          &$0$                & $0$    & XMM-Newton          & \citealt{Chen2018}        \\
                    & $5\,572$  & $5\,162$          &$0$                & $0$    & XMM-Newton          & \citealt{Chiappetti2013}  \\
                    %&$135\,118$ & $54$              &$20$               & $26$  & ROSAT                     &\citealt{Boller2016}    \\ %2RXS
                    %&$124\,730$ & $37$              &$8$                & $17$  & ROSAT                     &\citealt{Voges1999,Voges2000}    \\ %FSC(27,8,12) + 1RXS(10,0,5); total number: 105924 FSC, 18806 1RXS
                 %  &$510\,764$ & $2\,079$          &$316$              & $285$ &  -                        &\cite{Flesch2015}       \\ %Quasar   HMQ
   N$_{\rm sum}$    & -         & $10\,404$         &$4\,053$           &$2\,630$& -                         &  -  \\
        \hline %37+54+5242+5162=10495; 8+20+4053=4081; 17+26+2630=2673
        \hline
        \end{tabular}
        \begin{tablenotes}   
        \item[$^1$] The number of sources in the region is counted within the survey coverage, as shown in Eq.~\ref{eq:region}.
        \end{tablenotes}
        \end{threeparttable}
\label{tab:gc_catalogs}
\end{table*}

We cross-match detections with previously identified galaxy clusters with the offset $<3.0\arcmin$. 
In this process, we do not set the threshold of redshift differences, to avoid the inaccurate estimation of cluster redshift. 
The position of clusters, together with the X-ray contour and member galaxies, are overlaid over the $r$-band image. One example is shown in the second row of Fig.~\ref{fig:visual_check_example}.
By visual check, only clusters with positions matching well with the X-ray contour are taken as counterparts of our detections. 

\begin{figure*}
    \centering
    \includegraphics[width=0.95\textwidth]{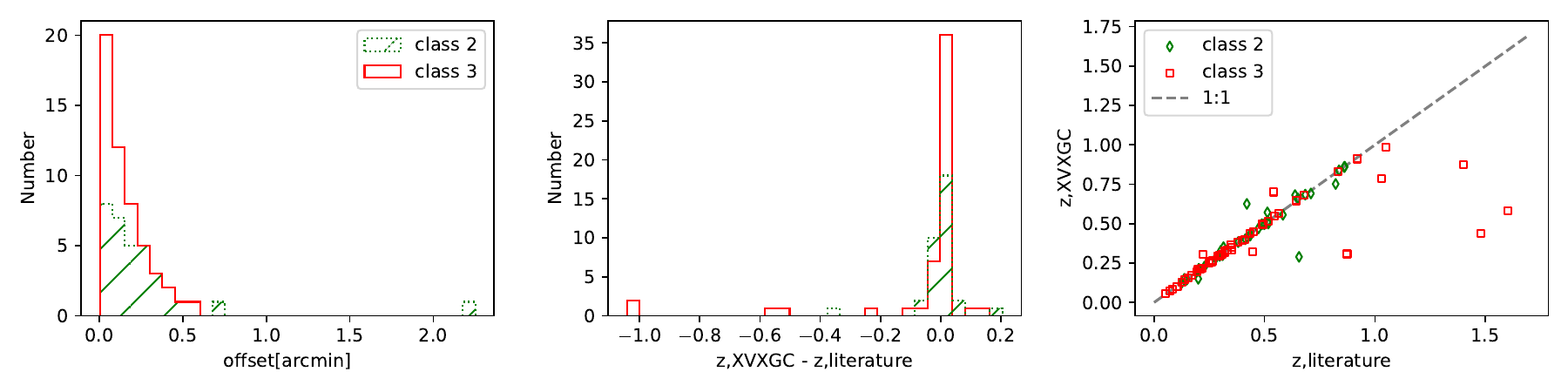}
    \caption{Left and the middle panel is the distribution of position offset and redshift difference for the cross-matched clusters between XVXGC and literature, respectively. The right panel is the redshift comparison. 
    The ``z,literature" in panels indicates the redshifts from the literature.
    Red solid histogram and red squares for class 3, green dot histogram and green diamonds for class 2. }
    \label{fig:hist_diff}
\end{figure*}

In Fig.~\ref{fig:hist_diff}, we show the distribution of position offsets and redshift differences between XVXGC clusters and previously detected cluster counterparts. 
Most matched clusters have an offset smaller than $0.5 \arcmin$. The redshift differences are always $<0.2$, except for 5 ICM-detected clusters, and 2 OPT/IR clusters.

% output of 41match_dis_deltaz.py for |delta_z|>0.2
%z_our,z_xsz: 0.8747 1.4
%z_our,z_xsz: 0.5821 1.6
%z_our,z_xsz: 0.7872 1.03
%z_our,z_xsz: 0.3079 0.874
%z_our,z_xsz: 0.4374 1.479
%z_our,z_opt: 0.6253 0.42
%z_our,z_opt: 0.2893 0.6562

With visual check, as shown in Fig.~\ref{fig:visual_check_example}, we classify our detections for genuine clusters into previous ICM-detected clusters (class 3), previous OPT/IR clusters (class 2), and new clusters (class 1). 
The detections with both previous ICM-detected clusters and OPT/IR clusters as counterparts are classified as class 3. 
The locations of these detections for genuine clusters are shown in Fig.~\ref{fig:radec}. 
In Fig.~\ref{fig:extent_extml}, their distribution in EXTENT-Extension likelihood parameter space is shown.
Finally, the numbers of clusters in classes are listed in Tab.~\ref{tab:n_class}. 

We compile the class 1, class 2, and class 3 detections into the final catalog, as described in Sec.~\ref{sec:result}. In later sections, we only discuss these cluster detections.

\begin{table}[]
    \centering
    \caption{Cluster numbers in different classes and regions.}
    \begin{tabular}{c|c | c c c  }
    \hline
    \hline
      & XMM-SERVS & XMM-LSS & W-CDF-S & ES1  \\
    \hline
    Total   & 141 &   80 & 37 & 24  \\
    \hline
    class 1 & 52  &   14 & 25 & 13  \\ 
    class 2 & 37  &   22 & 8  &  7  \\
    class 3 & 52  &   44 & 4  &  4  \\ 
    \hline
    \hline
    \end{tabular}
    \label{tab:n_class}
\end{table}

\section{Result and discussion}
\label{sec:result}

\subsection{The catalog}
\label{subsec:catalog}

In this work, we make detection of 141 X-ray extended galaxy clusters, named the XMM-SERVS X-ray eXtended Galaxy Cluster catalog (XVXGC). In Tab.~\ref{tab:xvxgc}, the first two entries are shown for the table format. 
The catalog and figures for each source, as shown in Fig.~\ref{fig:visual_check_example}, are available at the XVXGC webpage\footnote{\url{https://github.com/wwxu/xvxgc.github.io}} after the acceptance of this paper. 

\begin{table*}[]
    \caption{The first two entries in the XVXGC catalog.}
    %\tiny
    \centering
        \begin{threeparttable}
    \begin{tabular}{ c | c c c c c c c c c c c c }
\hline
\hline
 Name  & RAJ2000  & DEJ2000  & class  & N$_{\rm mem}$& $z$  & $z$,err& $z$,src& EXT$\_$ML & EXTENT  & $R_{\rm sig}$  & $CR_{\rm sig}$  & $CR_{\rm sig}$,err  \\
 -   & [deg]  & [deg]  & -  &  - &  -  &  - &  -&  - & [arcmin]  & [arcmin]  & [/s]  & [/s] \\
 (1) & (2)  & (3)  & (4) & (5)  & (6)   & (7)  & (8)  & (9)  & (10)  &(11)  & (12)  & (13)   \\
\hline
  es$\_$1 & 8.618 & -43.316 & 3 & 65 & 0.3945 & 0.0076 & zsp1 & 11951.611 & 0.526 & 3.307 & 0.082 & 0.001  \\ %es_706
  es$\_$2 & 8.801 & -43.170 & 1 & 14 & 0.1728 & 0.0063 & zsp1 &    24.410 & 0.112 & 2.555 & 0.008 & 0.001  \\  %es_535
\hline
\hline
\end{tabular}
\vspace{2mm}
    \begin{tabular}{c| c c c c c c c c c c c c c c}
\hline
\hline
  Name   & $R_{500}$  & $R_{500}$  & $R_{500}$,err &  $CR_{500}$  &  $CR_{500}$,err & $F_{500}$  & $F_{500}$,err  & $L_{500}$  & $L_{500}$,err  &  $T_{\rm x}$  & $T_{\rm x}$,err    \\
  & [arcmin]  & [Mpc]  & [Mpc] &  [/s]  &  [/s] &[erg/s/cm$^2$]  & [erg/s/cm$^2$]  &[erg/s]   & [erg/s]  & [keV]  & [keV]   \\
 (1)& (14) &(15) &(16) &  (17)  & (18)  & (19)  & (20)  &(21)  & (22)  & (23)  & (24)  \\
  \hline
 es$\_$1  & 2.611 & 0.835 & 0.001& 0.365 & 0.003& 2.937E-13 & 2.115E-15 & 1.605E44 & 1.156E42 & 4.315 & 0.009 \\
 es$\_$2  & 2.018 & 0.356 & 0.009& 0.013 & 0.002& 1.047E-14 & 1.489E-15 & 8.696E41 & 1.237E41 & 0.706 & 0.032 \\
\hline
\hline
\end{tabular}
\vspace{2mm}
    \begin{tabular}{c| c c c c c c c c c}
\hline
\hline
  Name  & $M_{500}$  & $M_{500}$,err  & $\beta$  & $\beta$,err1  & $\beta$,err2  & $R_{\rm c}$  & $R_{\rm c}$,err1  & $R_{\rm c}$,err2 &Comment \\
 -& [M$_{\odot}$]  & [M$_{\odot}$]& - & - &  - & [arcmin]  & [arcmin] & [arcmin] &-  \\
 (1)& (25)& (26) & (27)    & (28)  & (29)  & (30)  & (31)  &(32)&(33)\\
  \hline
 es$\_$1 & 2.506E14 & 8.447E11 & 0.683 & 0.004 & 0.004 & 0.663 & 0.007 & 0.007 &-\\
 es$\_$2 & 1.515E13 & 1.104E12 & 0.374 & 0.004 & 0.005 & 0.075 & 0.014 & 0.016 &-\\
\hline
\hline
\end{tabular}
        \begin{tablenotes}
        \item[\textsc{Notes.}] ``err" refers to the $1\sigma$ error, while ``err1" and ``err2" refer to the lower error and upper error in case of the asymmetric error. 
        \textsc{Column 8.} The redshift source, as described in Sec.~\ref{subsec:redshift_estimation}.
        \end{tablenotes}
        \end{threeparttable}
    \label{tab:xvxgc}
\end{table*}

\begin{figure*}
    \centering
    \includegraphics[width=0.95\textwidth]{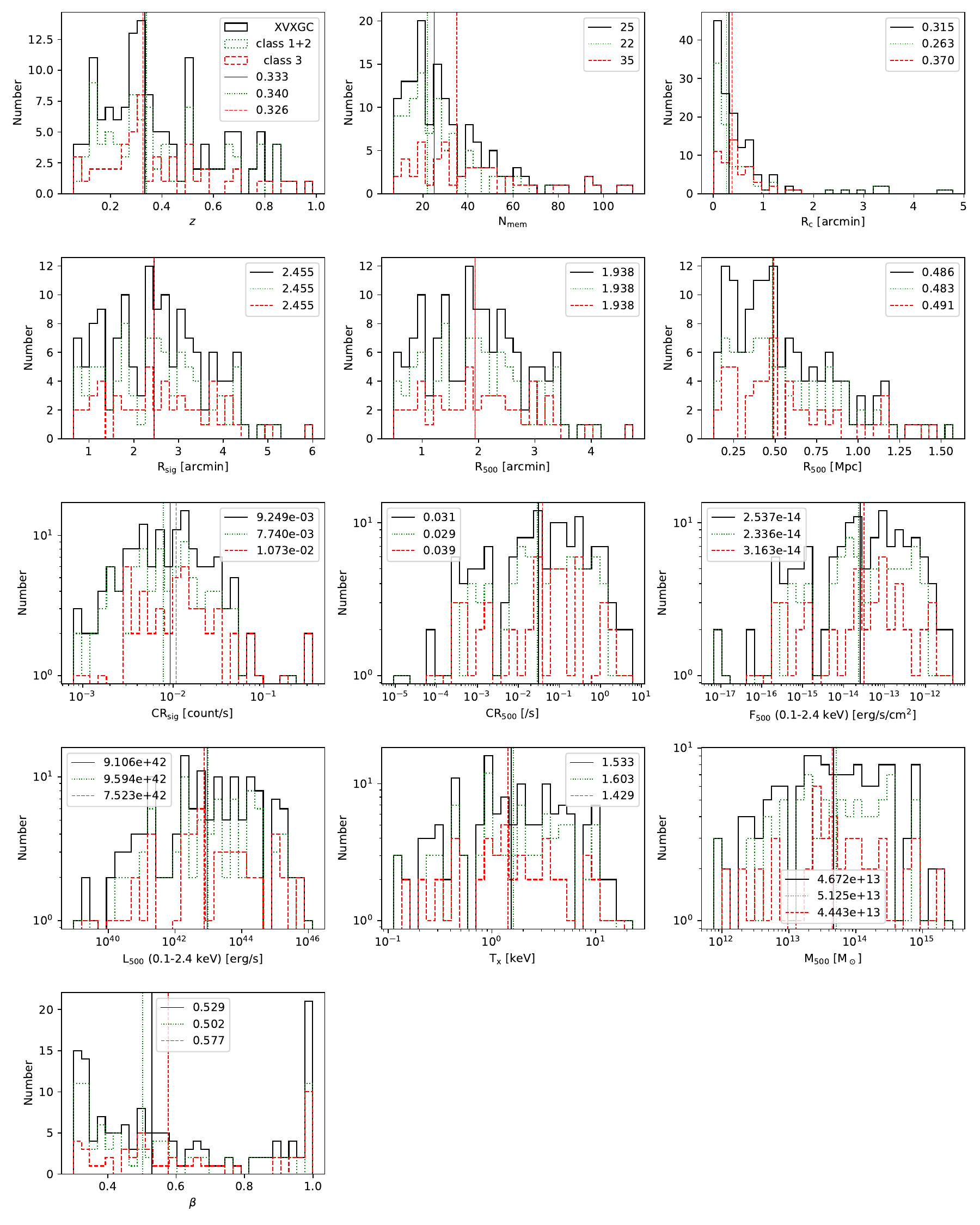}
    \caption{Histogram of parameters. The red dashed histogram for class 3 clusters, and green dotted histogram for the ``class 1 + class 2", and the black solid histogram for the whole XVXGC sample. The red dashed, green dotted, and black solid vertical line is the median value of the corresponding sample. }
    \label{fig:hist_para}
\end{figure*}

In the XVXGC catalog, there are new 52 clusters (class 1), 37 previous OPT/IR clusters (class 2), and 52 previous ICM-detected clusters (class 3). 
In Fig.~\ref{fig:hist_para}, we plot the histogram of parameters for the XVXGC sample, ``class 1 + class 2", and class 3 samples, with the median value overlaid. These median values are also listed in Tab.~\ref{tab:med_class}.
Besides, the median value of the $1\sigma$ error of $R_{500}$ is $0.005$~Mpc, $CR_{500}$ error as $0.003$/s, $F_{500}$ error as $2.115\times 10^{-15}$~erg/s/cm$^2$, $L_{500}$ error as $5.765\times 10^{41}$~erg/s, $T_{\rm x}$ error as $0.037$~keV, $M_{500}$ error as $1.892\times 10^{12}$~M$_{\odot}$.

%\subsection{Comparison class 3 with ``class 1 + class 2" clusters}
%In Fig.~\ref{fig:par_comparison}, the parameters are plotted with another. 
From this figure and the table, we find:
\begin{itemize}
    \item The redshifts of XVXGC sample spread from 0.005 to 1.0, and most detections are with $z<0.6$. 
    \item The number of members are in range of $7-113$, with median value of $25$.
    \item The median radius of XVXGC clusters is $\sim0.3\arcmin$, $\sim2.5\arcmin$, $\sim1.9\arcmin$, $\sim0.5$~Mpc for $R_{\rm c}$, $R_{\rm sig}$, $R_{500}$, and $R_{500}$.  
     \item The median count rate within $R_{\rm sig}$ and $R_{500}$ is $\sim0.01$ count/s and $\sim0.03$ count/s separately.
     \item The median flux, luminosity, gas temperature, and mass within $R_{500}$ is $2.5\times10^{-14}$~erg/s/cm$^2$, $9.1\times10^{42}$~erg/s, $1.5$~keV, $4.7\times10^{13}~M_{\odot}$. The histograms in the logarithm of these parameters are nearly symmetric. 
     \item A large fraction of clusters are with $M_{500}<10^{14}~M_{\odot}$, and should be classified as galaxy groups instead of galaxy clusters. However, in this work, we use the term ``galaxy clusters" representing both types for convenience.
     \item Except values at the high and low $\beta$ value end, representing ultra-steep or ultra-flat surface brightness profiles, the distribution of $\beta$-value peaks at the value of $\sim 0.5$. It corresponds to a much flatter surface brightness profile than the typical cluster profile.
     \item The class 1, 2, and 3 samples are found to overlap with each other in all parameters generally, although the ``class 1+2'' sample seems to always take a comparably larger percentage close to the lowest end. 
     \item Compared with class 3 clusters, the ``class 1 + class 2" sample tends to be less bright, less massive, with a flatter X-ray surface brightness profile.
\end{itemize}

\begin{table*}[]
    \centering
    \caption{Median value of parameters for the XVXGC sample, ``class 1 + class 2" sample, and class 3 sample.}
    \renewcommand\arraystretch{1.5}
    \begin{tabular}{c|c c c}
    \hline\hline
        Par. & Med. (XVXGC) & Med. (class 1+2) &  Med. (class 3) \\
    \hline
    $z$                 & 0.333$^{+ 0.013}_{- 0.015}$         & 0.340$^{+ 0.018}_{- 0.033}$         & 0.326$^{+ 0.022}_{- 0.020}$\\
Nmem                    & 25$^{+2}_{-1}$                      & 22$^{+  2}_{-  2}$                  & 35$^{+  7}_{-  5}$\\
$R_{\rm sig}$[arcmin]   & 2.455$^{+ 0.101}_{- 0.080}$         & 2.455$^{+ 0.101}_{- 0.147}$         & 2.455$^{+ 0.222}_{- 0.090}$\\
$R_{500}$[arcmin]       & 1.938$^{+ 0.080}_{- 0.063}$         & 1.938$^{+ 0.080}_{- 0.116}$         & 1.938$^{+ 0.175}_{- 0.071}$\\
$R_{500}$[Mpc]          & 0.486$^{+ 0.018}_{- 0.030}$         & 0.483$^{+ 0.052}_{- 0.042}$         & 0.491$^{+ 0.015}_{- 0.025}$\\
$R_{c}$[arcmin]         & 0.315$^{+ 0.048}_{- 0.024}$         & 0.263$^{+ 0.042}_{- 0.046}$         & 0.370$^{+ 0.069}_{- 0.015}$\\
$CR_{\rm sig}$[/s]      &9.249e-03$^{+1.175e-03}_{-1.472e-03}$&7.740e-03$^{+1.515e-03}_{-1.227e-03}$&1.073e-02$^{+2.424e-03}_{-6.715e-04}$\\
$CR_{500}$[/s]          & 0.031$^{+ 0.028}_{- 0.003}$         & 0.029$^{+ 0.030}_{- 0.007}$         & 0.039$^{+ 0.044}_{- 0.010}$\\
$F_{500}$[erg/s/cm$^2$] &2.537e-14$^{+2.264e-14}_{-3.572e-15}$&2.336e-14$^{+2.465e-14}_{-5.915e-15}$&3.163e-14$^{+3.710e-14}_{-7.982e-15}$\\
$L_{500}$[erg/s]        &9.106e+42$^{+7.154e+42}_{-3.429e+42}$&9.594e+42$^{+9.348e+42}_{-4.820e+42}$&7.523e+42$^{+1.217e+43}_{-1.928e+42}$\\
$T_{\rm x}$[keV]        & 1.533$^{+ 0.333}_{- 0.183}$         & 1.603$^{+ 0.399}_{- 0.387}$         & 1.429$^{+ 0.491}_{- 0.079}$\\
$M_{500}$[$M{_\odot}$]  &4.672e+13$^{+1.850e+13}_{-8.777e+12}$&5.125e+13$^{+1.744e+13}_{-1.425e+13}$&4.443e+13$^{+2.275e+13}_{-1.164e+13}$\\
$\beta$                 & 0.529$^{+ 0.026}_{- 0.027}$         & 0.502$^{+ 0.044}_{- 0.062}$         & 0.577$^{+ 0.059}_{- 0.049}$\\
    \hline\hline
    \end{tabular}
    \begin{tablenotes}
      \item[\textsc{Notes.}] The errors refer to  95$\%$ bootstrapped confidence interval for median value.
    \end{tablenotes}
    \label{tab:med_class}
\end{table*}

\subsection{Distribution of size estimations}

\begin{figure}
    \centering
    \includegraphics[width=0.45\textwidth]{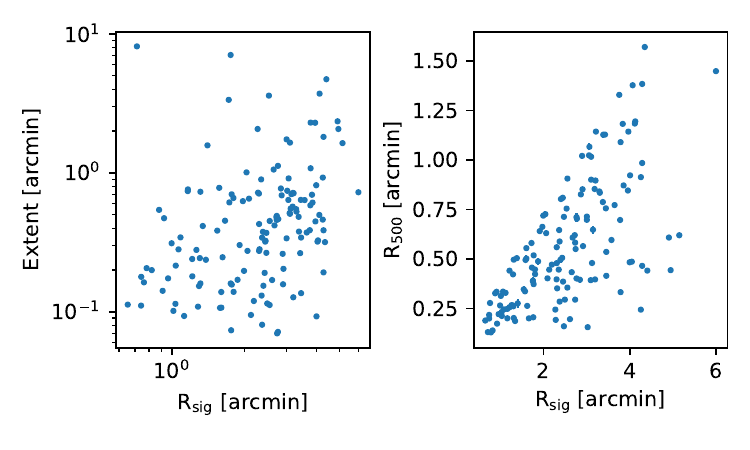}
    \caption{The distribution of radius estimations,  $R_{\rm sig}$, EXTENT, and $R_{500}$. The significant radius, $R_{\rm sig}$, is the radius when the growth curve turns into a plateau (Sec.~\ref{subsec:growth_curve}). The EXTENT is the estimation of the core radius in the $\beta$-model using the maximum likelihood fitting (Sec.~\ref{sec:detection}). The $R_{500}$ is the radius where the average density inside is $500$ times the critical density (Sec.~\ref{sec:flm}).}
    \label{fig:r_cr}
\end{figure}

In Fig.~\ref{fig:r_cr}, we show the distribution of size estimations. 
The significant radius is the outer boundary when the X-ray emission from the cluster is higher than the background fluctuation. The EXTENT value is estimated from the maximum likelihood fitting method, as a characteristic radius for extended sources. It is comparable with the significant radius and has a comparable weak correlation with the significant radius. However, the EXTENT parameter is only used in the separation of extended and point-like sources, and the threshold of EXTENT value ($4\arcsec$) is small for the detection of a large fraction of extended sources.

In addition, the $R_{500}$ is the radius when the average density within is 500 times the average cosmological density at the redshift. In the Figure, we find the $R_{500}$ is always lower than the significant radius.
It means in the analysis, all X-ray emission in the region of $R_{500}$ are considered.

\subsection{Flux function, luminosity function, mass function.}

\begin{figure}
    \centering
    \includegraphics[width=0.45\textwidth]{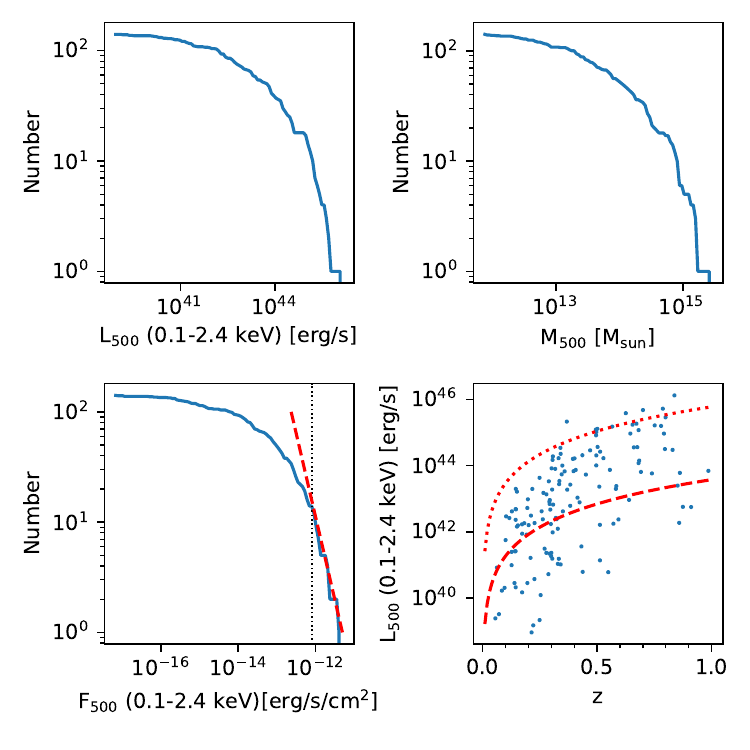}
    \caption{The cumulative number of luminosity, mass, and flux, as well as the relation of luminosity and redshift. In the lower-left panel, the dashed line indicates the line with the slope of $-1.5$ and normalized at $F_{500}=8 \times 10^{-13}$~erg/s/cm$^2$. In the lower-right panel, the dashed and dotted curve corresponds to the $F_{500}$ of $5 \times 10^{-15}$~erg/s/cm$^2$, and $8 \times 10^{-13}$~erg/s/cm$^2$, respectively.}
    \label{fig:fFunction}
\end{figure}

In the top-left, top-right, and lower-left panels of Fig.~\ref{fig:fFunction}, the cumulative number of luminosity, mass, and flux are shown as a representation for the luminosity, mass, and flux function, although no selection functions are corrected here. 
In the lower-left panel, the dashed line indicates the slope of $-1.5$ for the Euclidean universe when clusters are uniformly distributed. The line is normalized with the curve at $8 \times 10^{-13}$~erg/s/cm$^2$. It matches well with the curve at 
$>8 \times 10^{-13}$~erg/s/cm$^2$, which is shown with the vertical line. This indicates our high completeness for brighter clusters. 
In the histogram of $F_{500}$ (Fig.~\ref{fig:hist_para}), there are some ``class 1+2" clusters with $>8 \times 10^{-13}$~erg/s/cm$^2$. The newly ICM-detected bright clusters is a possible evidence of unexpected incompleteness in previous works.

In addition, the flux of $5 \times 10^{-15}$~erg/s/cm$^2$, and $8 \times 10^{-13}$~erg/s/cm$^2$ are overlaid into the relation of luminosity and redshift (lower-right panel of Fig.~\ref{fig:fFunction}). We find XVXGC clusters are mainly within this flux range up to the redshift of $\sim 1.0$, with some fainter clusters detected at low redshift of $z<0.5$.

\subsection{The offset to the X-ray sources}

\begin{figure}
    \centering
    \includegraphics[width=0.45\textwidth]{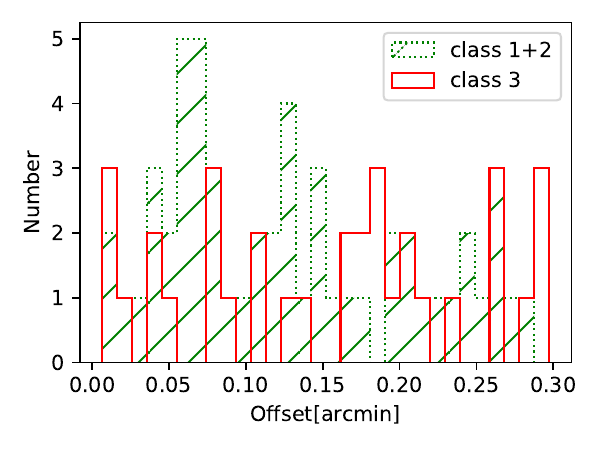}
    \caption{The histogram of offset between XVXGC clusters and X-ray sources. The red solid histogram is for class 3 clusters, and the green dotted histogram is for ``class 1 + class 2" clusters.}
    \label{fig:hist_xsrc}
\end{figure}

To quantify the effect of X-ray sources on our detections, we cross-match the XVXGC clusters with X-ray sources identified from ROSAT and XMM-Newton data, as listed in the last part of Tab.~\ref{tab:gc_catalogs}. The offset threshold is set as $0.315\arcmin$, which is the median value of the core radius of XVXGC clusters (Tab.~\ref{tab:med_class}).
%\sout{1.5\arcmin. This value is large enough for the size of the cluster's central area.}
There are $54$ ``class 1 + class 2" and $34$ `class 3' XVXGC clusters with central X-ray sources, and the offset distribution is shown in Fig.~\ref{fig:hist_xsrc}. 
% 18+71=89

\begin{figure*}
    \centering
    \includegraphics[width=0.9\textwidth]{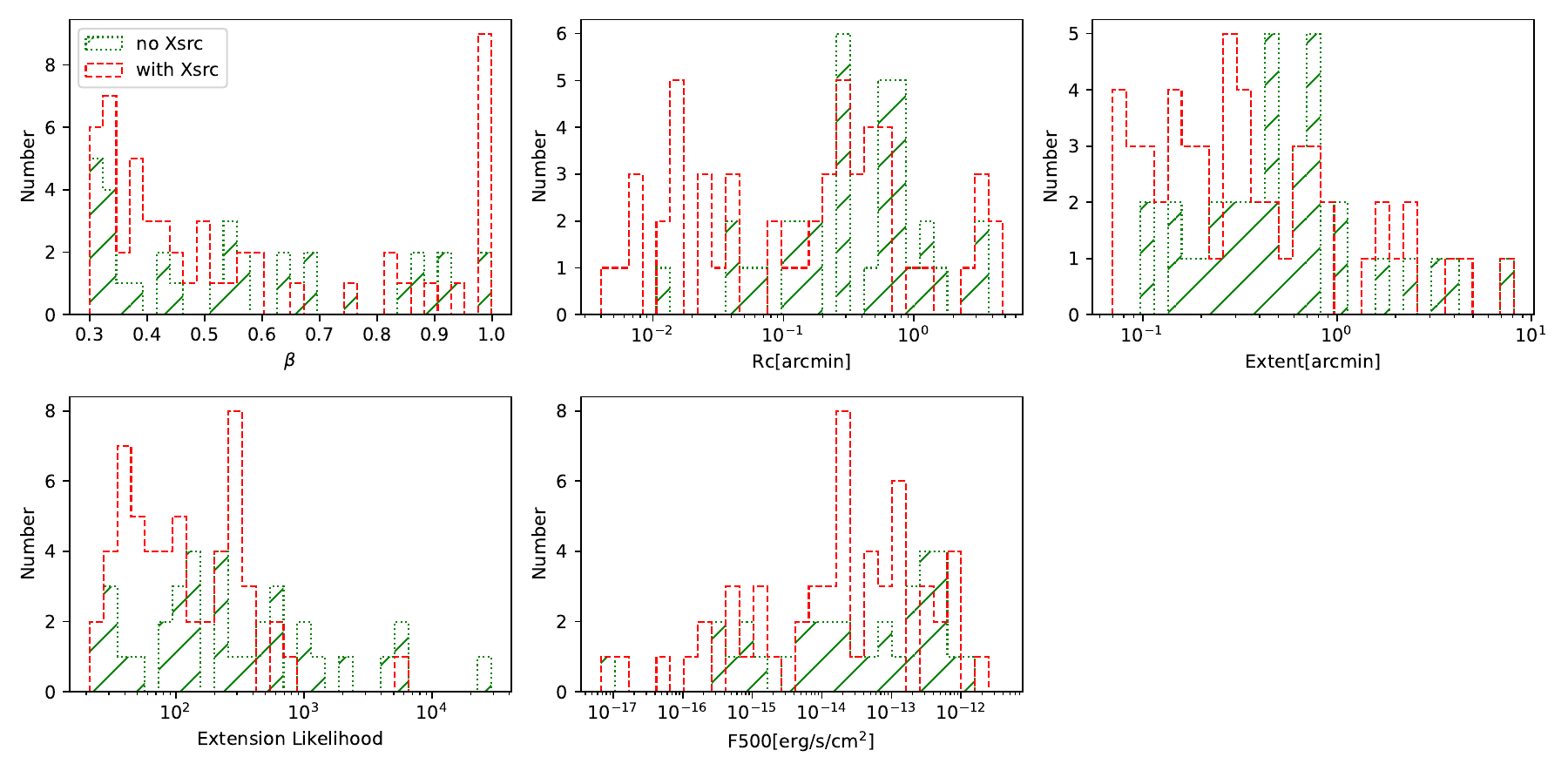}
    \caption{The parameter comparison for ``class 1+2" sample with and without central X-ray sources detected. }
    \label{fig:para_xsrc}
\end{figure*}

In the work of \cite{Ni2021} and \cite{Chen2018}, the average number density of X-ray sources identified in the XMM-SERVS survey is $\sim 0.25$ in each square arcmin. And, the total area of $141$ circles with a radius of $0.315\arcmin$ is $\sim 44$ square arcmin. Thus, there are $\sim 11$ X-ray sources within offset $<0.315\arcmin$ of XVXGC clusters from purely projection effect, assuming the X-ray sources are distributed randomly and without correlation with clusters.
It means there are roughly $46$ newly ICM-detected clusters without central X-ray sources, the percentage is $\sim 51\%$.
In Fig.~\ref{fig:para_xsrc}, the parameters are compared for the ``class 1+2" sample with and without central X-ray sources, and no systematic differences are found between the samples.
% 3.14*0.315*0.315*141 =43.930876500000004
% (4053+2630+5242)/(5.3+4.6+3.2)/3600.0 = 0.25286259541984735

\subsection{Physical parameter comparison with previous X-ray clusters}

\begin{figure*}
    \centering
    \includegraphics[width=0.95\textwidth]{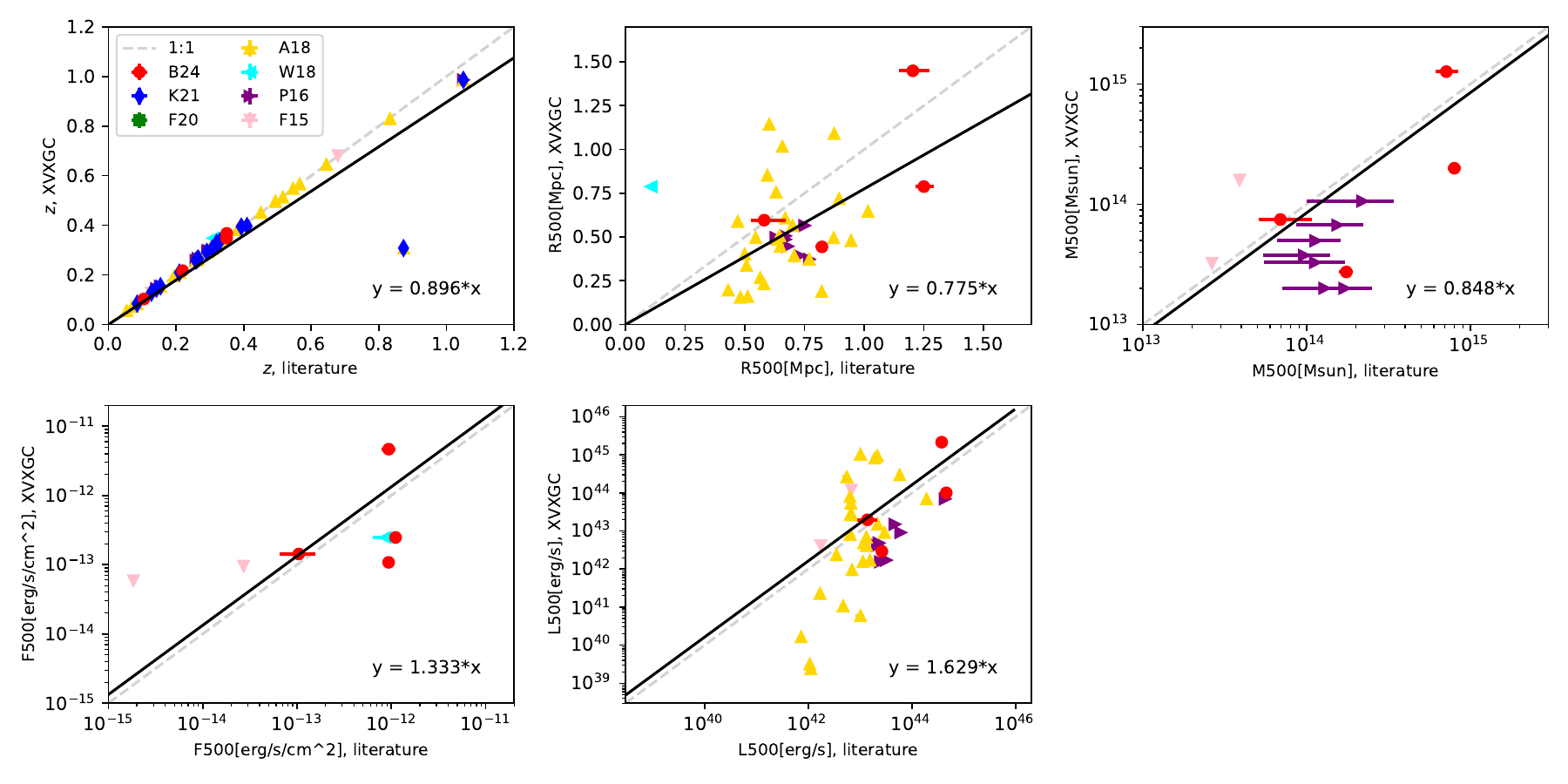}
    \caption{Comparison of physical parameters in XVXGC catalog and other X-ray cluster catalogs. The grey dashed lines show the relation of $y=x$. The best-fitting linear function is shown in a black line, with the equation written at the lower-right corner of each panel.}
    \label{fig:par_XrayGC}
\end{figure*}

We cross-match the XVXGC clusters with X-ray clusters from the literature, including those identified in \cite{Bulbul2024, Koulouridis2021, Finoguenov2020, Adami2018, Wen2018, Pierre2016, Finoguenov2015}. The cross-matching criteria is set as offset smaller than $1.0$~arcmin for a robust result. 
In XMM-LSS, W-CDF-S, ES1 region, there are respectively $30$, $4$, $4$ XVXGC clusters with X-ray clusters detected in these literature works.
The comparison of physical parameters is shown in Fig.~\ref{fig:par_XrayGC}. 
In general, parameter estimation is consistent systematically with these previous works. The estimation of redshift is tightly consistent with the literature values, although a few outliers exist.

Specifically, there are four XVXGC clusters detected in the primary eRASS1 cluster catalog \citep{Bulbul2024}. These four clusters are shown with red dots in Fig.~\ref{fig:par_XrayGC}. We find our estimation has no systematical bias, although some scatters exist.

%xxxx KS test
%z, KstestResult(statistic=0.04918032786885246, pvalue=0.9999997788048505, statistic_location=0.3079, statistic_sign=-1)
%R, KstestResult(statistic=0.47368421052631576, pvalue=0.0003162842604571903, statistic_location=0.505962, statistic_sign=-1)
%F, KstestResult(statistic=0.42857142857142855, pvalue=0.5751748251748252, statistic_location=2.473832e-13, statistic_sign=-1)
%L, KstestResult(statistic=0.28205128205128205, pvalue=0.08973505796946307, statistic_location=4.843888e+42, statistic_sign=-1)
%M, KstestResult(statistic=0.46153846153846156, pvalue=0.12648770263254042, statistic_location=67182840000000.0, statistic_sign=-1)

\section{Conclusion}
\label{sec:conclusion}

In this work, we make the detection of X-ray extended clusters based on the XMM-SERVS data \citep{Chen2018, Ni2021} with the wavelet-based detection method \citep{Pacaud2006, Xu2018}. We identified 141 clusters, named the XMM-SERVS X-ray eXtended Galaxy Cluster (XVXGC) catalog. There are 52 new clusters, 37 previously identified OPT/IR clusters without previous ICM-based identification, and 52 previous ICM-identified clusters. Compared with previous ICM-detected clusters, our newly ICM-detected clusters tend to be fainter, with a flatter X-ray surface brightness profile. 

Out of 89 first ICM-detected clusters, only $\sim 49\%$ have X-ray sources identified previously in the central region. Thus, there are roughly $46$ clusters detected with their ICM emission for the first time. In addition, by comparing physical parameters with counterparts in previous X-ray cluster catalogs, we find XVXGC has a good redshift and physical parameter estimations. 

\begin{acknowledgements} 
We acknowledge support from the National Key R\&D Program of China (2021YFA1600404, 2022YFF0503403, 2022YFF0503401), National Nature Science Foundation of China (Nos 11988101, 12022306, 12203063), the China Manned Space Project (CMS-CSST-2021-B01, CMS-CSST-2021-A01, CMS-CSST-2021-A05, CMS-CSST-2021-A06, CMS-CSST-2021-A07), and the National Science Foundation of China (12225301), the support from the Ministry of Science and Technology of China (Nos. 2020SKA0110100), CAS Project for Young Scientists in Basic Research (No. YSBR-062), and the support from K.C.Wong Education Foundation.
BL acknowledges financial support from the National Natural Science
Foundation of China grant 11991053.
CZ acknowledges support from Zhejiang Provincial Natural Science Foundation of China under grant No. LQ24A030001. 
WX thanks Thomas H. Reiprich, Florian Pacaud, Miriam E. Ramos-Ceja, Mingyang Zhuang, Shi Li for useful discussion during the development of this paper. 
% http://vipers.inaf.it/rel-pdr2.html
This paper uses data from the VIMOS Public Extragalactic Redshift Survey (VIPERS). VIPERS has been performed using the ESO Very Large Telescope, under the "Large Programme" 182.A-0886. The participating institutions and funding agencies are listed at http://vipers.inaf.it. 

\end{acknowledgements}

\bibliographystyle{aa} 
\bibliography{main} 

\end{document}